\documentclass[aps,prd,twocolumn,a4paper,nofootinbib,superscriptaddress,longbibliography]{revtex4-2}

\usepackage{amsmath,amssymb,bm,mathrsfs}
\usepackage{booktabs}
\usepackage{array}
\usepackage{slashed}
\usepackage[dvipdfmx]{graphicx}
\usepackage[dvipdfmx,colorlinks=true,linkcolor=blue,citecolor=blue,urlcolor=blue]{hyperref}
\usepackage{url}
\usepackage[dvipsnames]{xcolor}
\begin{document}

\title{Confinement--Higgs Complementarity \\ Viewed through Relative Quantum Reference Frames}

\author{Kei-Ichi Kondo}
\email{kondok@faculty.chiba-u.jp}
\affiliation{Department of Physics, Graduate School of Science, Chiba University, 1-33 Yayoi-cho, Chiba, Chiba 263-8522, Japan}
\affiliation{Research and Education Center for Natural Sciences, Keio University, 4-1-1 Hiyoshi, Yokohama, Kanagawa 223-8521, Japan}
\date{November 15, 2026; CHIBA-EP-285v1}

\begin{abstract}
We formulate confinement--Higgs complementarity in terms of the relative orientation between two independently constructed quantum reference frames (QRFs): one derived from the gauge field and the other from the scalar field.
Exact continuum and lattice identities give a relational interpretation of the scalar--color observables used in recent gauge-fixing-free $SU(2)$ simulations.
In particular,
we provide a QRF interpretation of the additional gauge-invariant nonlocal separation line obtained from the Cho--Duan--Ge (CDG) reduction reported in lattice simulations of the radially fixed fundamental $SU(2)$ gauge--scalar model.
We explain why such a candidate nonlocal boundary is compatible with Fradkin--Shenker continuity and Osterwalder--Seiler analyticity for the appropriate class of local gauge-invariant observables.
Separately, we show that a Wilson-rooted control observable admits an exact  results at zero plaquette coupling and that
it does not cause such a nonlocal separation line.

\end{abstract}

\maketitle
\tableofcontents

\section{Introduction}
\label{sec:intro}

The relation between color confinement \cite{Wilson1974,Polyakov1977} and the Brout--Englert--Higgs (BEH) phenomenon \cite{EnglertBrout1964,Higgs1964,GuralnikHagenKibble1964,
Higgs1966,Kibble1967} is one of the places where the use of gauge-dependent language can obscure the physical question.  In a gauge-fixed perturbative description the Higgs regime is commonly represented by a scalar vacuum expectation value and massive gauge fields, whereas a confinement regime is described in terms of gauge-invariant bound states, flux tubes, screening, and the absence of isolated charged asymptotic sectors.  Comparing these two descriptions literally can therefore compare unlike objects.  The older idea of confinement--Higgs complementarity already points toward a gauge-invariant formulation of the problem.

The Wilson loop criterion \cite{Wilson1974} supplies an early gauge-invariant confinement diagnostic: the area-law behavior of a test Wilson loop at intermediate or asymptotic distances must be distinguished from the screening behavior in the presence of dynamical fundamental matter.

't Hooft \cite{tHooft1978} emphasized that confinement and Higgs phenomena should be classified by the algebra of physical operators, order--disorder variables, and electric--magnetic structure, rather than by a gauge-dependent order parameter alone.
Osterwalder and Seiler \cite{OsterwalderSeiler} established rigorous analyticity results for lattice gauge theories and a nonperturbative treatment of the Higgs mechanism.
Fradkin and Shenker \cite{FradkinShenker} showed that, for \textit{fixed-length} scalar matter in the fundamental representation, there exists an analytically connected region (and paths within it) joining confinement-like and Higgs-like regimes for the class of local gauge-invariant observables covered by their theorem.
For nonfundamental matter, including the $SU(2)$ adjoint representation, phase boundaries can occur.
In this analyticity statement, locality is crucial: the relevant observables have support that remains contained in a compact set in the thermodynamic limit, whereas a nonlocal observable may display nonanalytic behavior without contradicting the local Osterwalder--Seiler--Fradkin--Shenker (OSFS) analyticity domain \cite{Bonati2009}.
The Fr\"ohlich--Morchio--Strocchi (FMS) construction \cite{FMS1980,FMS1981} subsequently made the gauge-invariant spectrum central: a gauge-invariant composite may interpolate the same physical state whose leading gauge-fixed expansion contains an elementary Higgs or vector field.
These developments make complementarity a statement about gauge-invariant physics, not a statement that all dynamical characteristics of the two regimes are identical.

This distinction has become especially important in modern lattice studies.
\textbf{Nonlocal observables can distinguish regimes even when all local gauge-invariant thermodynamic functions remain analytic.}
Greensite and Matsuyama  \cite{GreensiteMatsuyama2017,GreensiteMatsuyama2018,
GreensiteMatsuyama2022,GreensiteYau2024} introduced a gauge-invariant but nonlocal diagnostic of a global custodial symmetry and argued for a distinction between ordinary color neutrality and a stronger separation-of-charge notion of confinement.  Because their diagnostic is nonlocal, a nonanalyticity of that observable need not contradict Osterwalder--Seiler--Fradkin--Shenker (OSFS) analyticity.  Related lattice work has studied string breaking, vortices, monopoles, and nonlocal order/disorder observables in gauge--Higgs systems \cite{KnechtliSommer,GreensiteOlejnik,Wenzel2005,ShimadaYamamoto2025}.

The present article is built on two earlier QRF formulations for non-Abelian gauge theories.  Ref.~\cite{KondoQRFConfinement2026} formulated color confinement in terms of relational charged sectors, Gauss's law, and the availability or failure of admissible long-distance color frames.  Ref.~\cite{KondoQRFBEH2026} formulated the BEH mechanism by treating the Higgs orientation as an internal QRF and the massive vector as the gauge connection seen relative to that dynamical frame.  We regard the detailed constructions in those two works as established input and do not reproduce them in full.
\textbf{The new problem is to understand what happens when both kinds of reference structure are present in the same gauge--matter theory.}

The key proposal is simple.  Let $h_{\mathscr A}(x)=h[\mathscr{A}](x)$ denote a confinement-oriented frame constructed from the gauge field $\mathscr{A}$, for example by a Wilson-line prescription or by a gauge-covariant color-direction construction \cite{Kondo:2014sta}, and let $h_{\Phi}(x)=h[\Phi](x)$ denote a Higgs-oriented frame constructed from scalar matter $\Phi$.
For strictly left-covariant frames, they have the left local gauge transformation:
$h_{\mathscr A}(x)\mapsto\Omega(x) h_{\mathscr A}(x)$ and $h_{\Phi}(x)\mapsto\Omega(x) h_{\Phi}(x)$,
Then the relative frame
\begin{equation}
 q(x)=h_{\Phi}[\Phi](x)^{\dagger}h_{\mathscr A}[\mathscr{A}](x)
 \label{eq:qintro}
\end{equation}
is pointwise gauge invariant $q(x)\mapsto q(x)$ for strictly left-covariant frames.
For a Wilson-rooted frame $h_{\mathscr A}$ the transformation includes a right action at the common root $x_0$, as detailed in Sec.~\ref{sec:continuumgeneral}; correspondingly, $q(x)\mapsto q(x)\Omega(x_0)^{-1}$.
Traced relative two-point functions are invariant. Hence $q$ represents a relative orientation between two frames rather than, by itself, a physical gauge transformation.
Its physical interpretation always includes the frame/root prescription.

\textbf{The distinction between confinement-like and Higgs-like behavior can then be encoded in the long-distance coherence, locking, defects, and residual stabilizer of this relative frame.}
 Importantly, the \emph{existence} of $q$ at a finite-volume configuration is generally too weak to characterize a phase.
\textbf{The relevant questions are whether the two frames can be defined with an admissible prescription, whether the relative data are local or nonlocal, whether they possess a finite correlation length, whether zeros or defects obstruct globalization, and whether the corresponding charged sectors have finite or divergent isolation energy.}

This viewpoint is directly relevant to two gauge-independent lattice simulations \cite{Ikeda2024,ShibataKondo2024}.
Ikeda, Kato, Kondo, and Shibata \cite{Ikeda2024} studied the $SU(2)$ gauge--scalar model with a \textit{radially fixed} \textbf{fundamental scalar}. They reproduced the conventional weak-coupling thermodynamic transition line and reported, \textbf{at strong gauge coupling, an additional gauge-invariant nonlocal separation line detected by operators built from the fundamental scalar $\Phi$ and a gauge-field-derived color-direction field $\bm n_\mathscr{A}$}.
We refer to the latter below as a reported or candidate \textbf{relative-frame boundary} unless its infinite-volume critical character has been established independently.
\textbf{Its existence is compatible with OSFS analyticity because it need not define a universal local thermodynamic phase boundary.}
Shibata and Kondo \cite{ShibataKondo2024} performed the analogous gauge-fixing-free study for the $SU(2)$ gauge--scalar model with a \textit{radially fixed} \textbf{adjoint scalar}, reproduced the known confinement--Higgs boundary, and reported \textbf{an additional boundary dividing the confinement region into two parts}.
Further numerical evidence in the fundamental $SU(2)$ lattice model was reported by Shibata and Kondo \cite{ShibataKondo2025} following Ref.~\cite{Ikeda2024}; this later study examines the reported separation line and monopole observables.
Neither work is, by itself, a determination of universal continuum critical behavior.

\textbf{Our primary purpose is to propose and test a QRF interpretation of these numerical results:
the fundamental observable is read as a complete matter frame relative to a gauge-derived partial color frame, while the adjoint observable compares two partial frames and leaves a residual compact $U(1)$ sector whose infrared dynamics must be tested independently.}

There is also an important continuum question hidden by the \textit{fixed-length} assumption.  When the scalar radial amplitude fluctuates, the normalized scalar orientation fails at scalar zeros.  Consequently, a Higgs QRF is only patchwise defined and may undergo frame slips or acquire defect structure, even if the unnormalized scalar field is perfectly regular.
We therefore first formulate the fundamental theory
with their radial modes in the continuum, and only afterward pass to the lattice.  This ordering makes clear which statements are kinematical identities, which are dynamical approximations, and which are special consequences of a fixed-length model.

The relation to earlier gauge-invariant constructions is important for fixing the scope of the present proposal. Gauge-invariant dressings and boundary frames, the dressing-field method, and gauge-invariant treatments of the Higgs sector already show how gauge redundancy can be traded for composite relational variables \cite{GomesHopfmullerRiello,BorsboomPosthuma2026,AttardEtAl,MassonWallet,LavelleMcMullan,IldertonLavelleMcMullan}. The FMS construction likewise makes the gauge-invariant composite spectrum central \cite{FMS1980,FMS1981}.

The new operational proposal is the \emph{same-coupling comparison} of two independently prescribed dynamical frames, one gauge-derived and one matter-derived, through stabilizer-invariant relational observables and matched tests of charged-sector, defect, and gauge-invariant spectral diagnostics. This is neither a new dressing-field identity, nor a Hilbert-space frame-change unitary, nor a Gribov-free quantization.

A Wilson-rooted frame will be retained below only as an exactly specified control prescription. Its two-point relative-frame correlator is exactly a prescribed-path scalar--string correlator and is not the same operator as the Cho--Duan--Ge (CDG) relative-color observable \cite{Cho:1980,DuanGe:1979,KatoKondoEtAl2006,Kondo:2014sta}. Consequently a Wilson-tree identity cannot by itself determine a CDG separation line or critical exponent; agreement or disagreement between the two prescriptions is instead part of the proposed falsification program.

A further motivation comes from extended objects.  Hayata, Hidaka, and Kondo  \cite{HayataHidakaKondo2025} found a genuine second-order transition on superfluid vortex worldsheets inside a bulk confinement--Higgs crossover: a $\mathbb{Z}_2$ flavor symmetry is broken on the vortex in the weak-coupling Higgs regime and restored in the strong-coupling confinement regime, with two-dimensional Ising universality.  We shall argue that this is naturally compatible with the QRF picture.  A bulk analyticity statement for local observables does not forbid a nonanalytic change in a defect-supported relational sector.  Closely related lessons arise from the smooth matching of Aharonov--Bohm phases across confinement--Higgs continuity \cite{Cherman2020,Hayashi2024}, from Kertesz lines associated with vortex proliferation \cite{Wenzel2005}, and from recent nonlocal-operator studies \cite{ShimadaYamamoto2025}.

The organization of this article is as follows.
Sec.~\ref{sec:continuumgeneral} develops the continuum relative-frame construction, states the minimal QRF meaning used here, and retains Wilson-rooted transport only as a control prescription before distinguishing a QRF change from a Fradkin--Shenker interpolation.
Sec.~\ref{sec:fundcont} treats the fundamental scalar, including a
fluctuating radial mode. Sec.~\ref{sec:lattice} gives the lattice
formulation and controlled limits, while Sec.~\ref{sec:interpret}
interprets the numerical results of Refs.~\cite{Ikeda2024,ShibataKondo2025}.
Sec.~\ref{sec:radialpred} gives predictions for an unfixed radial
mode, and Sec.~\ref{sec:defects} discusses extensions and
defect-supported distinctions.
The numerical and continuum consistency tests are summarized in the concluding discussion and technical appendices.
Sec.~\ref{sec:conclusion} summarizes the physical status of the proposal. Technical derivations, including the Wilson-rooted control construction and the worldline check, are collected in the Appendices.

\section{Continuum relative-frame formulation}
\label{sec:continuumgeneral}

\subsection{Two admissible frames and an exact relational identity}

Let the gauge group be $G=SU(2)$ with Hermitian generators $T^A=\sigma^A/2$ and $\operatorname{tr}(T^AT^B)=\delta^{AB}/2$.
We use the Hermitian realization of the gauge connection and the covariant derivative,
\begin{equation}
 \mathscr A_\mu(x)=\mathscr A_\mu^A(x)T^A,
 \qquad D_\mu[\mathscr A]=\partial_\mu-ig\mathscr A_\mu,
 \label{eq:AF}
\end{equation}
where all components $\mathscr A_\mu^A$ are real.
The gauge field (connection) transforms under a local gauge transformation as
\begin{equation}
 \mathscr{A}_{\mu}(x)\mapsto
 \Omega(x)\mathscr{A}_{\mu}(x)\Omega(x)^{-1}
 +\frac{i}{g}\Omega(x)\partial_{\mu}\Omega(x)^{-1}, \ \Omega(x)\in G,
 \label{eq:gtrans}
\end{equation}

We introduce a group-valued \textbf{frame} $h(x)$ that transforms under the left local gauge transformation $\Omega(x)\in G$ as
\begin{equation}
 h(x)\mapsto \Omega(x)h(x), \ \Omega(x)\in G.
 \label{eq:htrans}
\end{equation}
Then we define the \textbf{relational gauge field (relational connection)}
\begin{equation}
 \mathscr{A}^{h}_{\mu}(x)
 :=h(x)^{-1}\mathscr{A}_{\mu}(x)h(x)
 +\frac{i}{g}h(x)^{-1}\partial_{\mu}h(x).
 \label{eq:Ah}
\end{equation}
This connection is invariant $\mathscr{A}^{h}_{\mu}(x)\mapsto \mathscr{A}^{h}_{\mu}(x)$ under the local left gauge transformation
\emph{provided} that $h$ transforms strictly as in Eq.~\eqref{eq:htrans}.

For the field strength defined by
\begin{align}
 \mathscr{F}_{\mu\nu}[\mathscr{A}](x)&=
 \partial_\mu\mathscr{A}_\nu(x)-\partial_\nu\mathscr{A}_\mu(x)
 -ig[\mathscr{A}_\mu(x),\mathscr{A}_\nu(x)],
 \label{eq:Fdefinitionadd}
\end{align}
the \textbf{relational field strength (relational curvature)} on a smooth frame patch is defined by
\begin{align}
 \mathscr F^h_{\mu\nu}(x)
 =&\partial_\mu\mathscr A^h_\nu(x)
 -\partial_\nu\mathscr A^h_\mu(x)
 -ig[\mathscr A^h_\mu(x),\mathscr A^h_\nu(x)]
 \nonumber\\
 =&h(x)^{-1}\mathscr F_{\mu\nu}[\mathscr A](x)h(x)
 =\mathscr{F}_{\mu\nu}[\mathscr{A}^h](x).
 \label{eq:relcurvatureIII}
\end{align}
This curvature is invariant $\mathscr F^{h}_{\mu\nu}(x)\mapsto \mathscr F^{h}_{\mu\nu}(x)$ under the simultaneous local left gauge transformation.
The equality assumes differentiability of $h$ on the patch;
a singular frame requires patchwise expressions and transition functions.

A formula such as Eq.~\eqref{eq:Ah} does not by itself guarantee that $h$ is a globally defined, local, unique, or normalizable quantum observable. We call a frame \emph{admissible} only relative to a specified domain, boundary protocol, charge sector, and operator renormalization, following the distinction emphasized in Ref.~\cite{KondoQRFConfinement2026}.

Here the term ``QRF'' is used in a deliberately minimal field-theoretic sense: the internal frame data are dynamical and fluctuate in the path integral; gauge-covariant frames are used to define relational observables; and no independent operator-valued unitary frame-change theorem on Hilbert space is derived from the identities below. This usage is compatible with canonical QRF constructions and recent relational gauge-theory formulations \cite{deLaHametteGalley2020,AraujoRegadoHohnSartini2025,FewsterJanssenRejzner2025}. Local gauge redundancy is not treated as a spontaneously broken physical symmetry; any ordering statement below refers either to an actual global symmetry or to a specified nonlocal relational diagnostic, consistently with Elitzur's theorem \cite{Elitzur1975}.

We now choose two frames $h_{\mathscr A}(x)$ and $h_{\Phi}(x)$.
The first, $h_{\mathscr A}(x)=h[\mathscr{A}](x)$, is constructed from gauge-field $\mathscr A$ and is adapted to confinement diagnostics.  It may be a Wilson-line frame, a boundary-rooted dressing, a Laplacian frame, or a lifted color-direction frame.
The second, $h_{\Phi}(x)=h[\Phi](x)$, is constructed from the scalar field $\Phi$ and is adapted to the BEH mechanism.

When both are \textbf{complete frames}, the \textbf{relative frame} $q(x)$ defined by
\begin{equation}
 q(x)=h_{\Phi}(x)^{-1}h_{\mathscr A}(x)
 \label{eq:qdef}
\end{equation}
is invariant under local left gauge transformations $q(x)\mapsto q(x)$
when \emph{both} frames are strictly left covariant.

In the common-root Wilson prescription, however, $h_{\mathscr A}(x)\mapsto\Omega(x)h_{\mathscr A}(x)
\Omega(x_0)^{-1}$ and $h_\Phi(x)\mapsto\Omega(x) h_\Phi(x)$ imply
$q(x)\mapsto q(x)\Omega(x_0)^{-1}$: only root-invariant contractions are observables.
This residual root action must not be identified with the independent right custodial action of the fundamental scalar.

In either case the two connections obey the following exact change-of-frame relation:
\begin{equation}
\mathscr{A}^{h_{\Phi}}_{\mu}(x)
 =q(x)\mathscr{A}^{h_{\mathscr A}}_{\mu}(x)q(x)^{\dagger}
 +\frac{i}{g}q(x)\partial_{\mu}q(x)^{\dagger}.
 \label{eq:framechange}
\end{equation}
It should be remarked that no equation of motion, gauge condition, or perturbative expansion has entered this identity.

For \textbf{partial frames}, a choice of group-valued lift
is additionally defined only modulo the stabilizer of the coset.
For example, an adjoint color direction $\bm n(x)=h(x)\sigma_3h(x)^\dagger$
leaves $h(x)\to h(x)k(x)$, $k(x)\in U(1)_{\sigma_3}$,
undetermined. If both frames are partial, their relative lift obeys
\begin{equation}
q(x)\longmapsto k_\Phi(x)^{-1}
 q(x)k_{\mathscr A}(x).
 \label{eq:doublecosetIII}
\end{equation}
Thus a raw group-valued $q$ is not an observable of two coset
directions without stabilizer contractions or transporters.
The corresponding gauge-invariant data belong to a double-coset
(or to explicitly lifted, patch-dependent fields), while the residual
stabilizer connection and Gauss constraint remain dynamical.
Patch transitions and defects cannot be removed merely by a local
choice of lift.

\subsection{Wilson-rooted transport as a control prescription}
\label{subsec:wilsontreecontrol}

\textbf{A Wilson-rooted frame} is useful as an exactly specified {control prescription}, although it is not the defining CDG frame of the present proposal.

Let $T$ be a rooted spanning tree and $P_x=(x_0,x_1,\ldots,x_m=x)$
its oriented root-to-$x$ path. Define $U_{a,b}$ on a forward
edge $a\to b$ by
$U_{a,b}\mapsto\Omega_aU_{a,b}\Omega_b^\dagger$.
Our transport \emph{from the root to $x$ in the frame convention} is
\begin{align}
 h_{\mathscr A,T}(x)
 &:=U(P_x):=U_{x_{m-1},x_m}^\dagger\cdots
 U_{x_0,x_1}^\dagger,\nonumber\\
 h_{\mathscr A,T}(x)&\mapsto\Omega_x h_{\mathscr A,T}(x)
 \Omega_{x_0}^\dagger.\label{eq:rooted-order-A4}
\end{align}
Accordingly $U(T_{xy}):=h_{\mathscr A,T}(x)
 h_{\mathscr A,T}(y)^\dagger$ denotes the map from the fiber
at $y$ to that at $x$ and transforms as
$\Omega_xU(T_{xy})\Omega_y^\dagger$.

A transported color axis is covariant only when its root datum transforms with the common root fiber:
\begin{equation}
 \bm n_{\mathscr A,T}(x)=h_{\mathscr A,T}(x)\bm n_0h_{\mathscr A,T}(x)^{-1},
 \quad
 \bm n_0\mapsto\Omega(x_0)\bm n_0\Omega(x_0)^{-1} .
 \label{eq:rootcovaxis}
\end{equation}
For a complete scalar frame $h_\Phi$, the shared root segment cancels configuration by configuration and gives the exact identity
\begin{equation}
\frac12\operatorname{tr}\!\left[q_T(x)q_T(y)^\dagger\right]
 =\frac12\operatorname{tr}\!\left[
 h_\Phi(x)^\dagger U(T_{xy})h_\Phi(y)\right],
 \label{eq:propositiontree}
\end{equation}
where $T_{xy}$ is the unique tree path between $x$ and $y$. Thus \textbf{the Wilson-tree correlator} is precisely a prescribed-path \textbf{scalar--string correlator}, not an independent two-frame order parameter.
Its path length, perimeter/end-point renormalization, and tree distortion are part of the operator definition.

Most importantly, Eq.~\eqref{eq:propositiontree} neither requires nor establishes \textbf{the CDG reduction condition}.
A reduction-derived color direction is a different functional of the gauge field and its lift has a local stabilizer freedom. Therefore no equality of transition point, correlation length, or critical exponent between \textbf{the Wilson-tree correlator} and the \textbf{CDG relative-color correlator} follows algebraically.
In what follows the tree construction is retained only to make this distinction explicit and to provide a controlled comparison prescription. Further root-fiber geometry, the spanning-tree figure, and the radial Wilson-frame remarks are collected in
Appendix~\ref{app:wilsontree}.

\subsection{A QRF change is not a Fradkin--Shenker interpolation}
\label{subsec:qrfvsfs}

A change from $h_{\mathscr A}$ to $h_{\Phi}$ compares the \emph{same configuration or state at the same couplings} from two different relational frames.  By contrast, a Fradkin--Shenker path changes the coupling constants,
\begin{equation}
 (\beta,\kappa,\lambda)\longrightarrow
 (\beta',\kappa',\lambda'),
\label{eq:FSpath}
\end{equation}
and follows a family of vacuum states through theory space.
\begin{equation*}
 \begin{split}
 &\text{confinement--Higgs complementarity}\\
 \neq &\text{a QRF transformation}.
 \end{split}
\end{equation*}
The QRF formulation instead explains how the physical degrees of freedom along such a path can be compared using gauge-invariant relational variables.

The point matters because a continuous Fradkin--Shenker path allows masses, screening lengths, defect densities, frame-correlation lengths, and bound-state sizes to vary smoothly, possibly rapidly.
The theorem says that \textbf{the relevant local gauge-invariant observables need not become nonanalytic; it does not say that every nonlocal or relational structure is constant.}
 Consequently, a QRF-regime boundary may be

\noindent
 (a) a \textbf{thermodynamic phase transition},

\noindent
 (b) a \textbf{nonlocal transition} not visible in the local free-energy density, or

\noindent
 (c) a \textbf{crossover}.

These possibilities must be distinguished by finite-size scaling rather than terminology.

\subsection{Relative-frame and its coherence}

To investigate the candidate nonlocal transition line, the lattice studies \cite{Ikeda2024,ShibataKondo2025} investigated the following $su(2)$ Lie algebra-valued field operator in the continuum notation:
\begin{align}
 r(x)&:=\hat\Theta(x)^\dagger \bm n_{\mathscr{A}}(x)\hat\Theta(x) \in su(2),
\end{align}
where $\bm n_{\mathscr{A}}(x)=\bm n[\mathscr A](x)$ is the CDG color-direction  field.
The CDG color-direction  field $\bm n(x)$ is written in the following form (adjoint-orbit representation) using a SU(2) group-valued field $h_{\mathscr{A}}(x) \in SU(2)$:
\begin{equation}
\bm n_{\mathscr{A}}(x)=h_{\mathscr{A}}(x)\sigma_3h_{\mathscr{A}}(x)^\dagger \in su(2), \quad  h_{\mathscr{A}}(x) \in SU(2) .
\end{equation}
This leads to another expression of $r(x) \in su(2)$:
\begin{align}
 r(x) = q(x) \sigma_3 q(x)^\dagger \in su(2),
\end{align}
which follows from
\begin{align}
r(x)=&\hat\Theta(x)^{\dagger} \bm n_{\mathscr{A}}(x) \hat\Theta(x)
= \hat\Theta(x)^{\dagger} h_{\mathscr{A}}(x) \sigma_3 h_{\mathscr{A}}(x)^\dagger \hat\Theta(x) \nonumber\\
 =& h_{\Phi}(x)^\dagger h_{\mathscr{A}}(x) \sigma_3 h_{\mathscr{A}}(x)^\dagger h_{\Phi}(x)
 = q(x) \sigma_3 q(x)^\dagger .
\end{align}
Now we find that \textbf{the scalar--color operator $r(x)$} introduced in Refs.~\cite{Ikeda2024,ShibataKondo2025} \textbf{has the QRF interpretation as the relative frame $q(x)=h_\Phi(x)^\dagger h_{\mathscr A}(x)$, namely the gauge frame measured in the scalar frame $h_\Phi(x)=\hat\Theta(x)$}:
\begin{equation}
q(x)
= h_{\Phi}(x)^\dagger h_{\mathscr{A}}(x) \in SU(2).
\end{equation}
To test this QRF interpretation, we propose to measure the \textbf{coherence of the relative-frame $q$}.
In what follows, we consider which type of correlator can be used for this purpose.

First, we define \textbf{the path-transported scalar correlator} for two complete frames
\begin{equation}
 \mathcal{C}_{\rm rel}(x,y;\Gamma)
 =\frac12\left\langle\operatorname{tr}\left[
 q(x)\mathcal{U}_{\mathscr{A}}(x,y;\Gamma)q(y)^\dagger
 \right]\right\rangle ,
 \label{eq:Crel}
\end{equation}
with \textbf{the confinement-frame transporter}
\begin{equation}
 \mathcal{U}_{\mathscr{A}}(x,y;\Gamma):=
 h_{\mathscr A}^\dagger(x)\,\mathcal{P}\exp\!\left(ig\int_{\Gamma:y\to x}dz^\mu\mathscr{A}_\mu(z) \right)h_{\mathscr A}(y).
 \label{eq:UC}
\end{equation}
Substitution of $q=h_{\Phi}^\dagger h_{\mathscr A}$ shows an exact cancellation at each field configuration:
\begin{align}
 q(x)\mathcal U_{\mathscr{A}}(x,y;\Gamma)q(y)^\dagger
 &=h_{\Phi}(x)^\dagger h_{\mathscr A}(x)h_{\mathscr A}(x)^\dagger \,U_\Gamma(x,y)\nonumber\\
 &\quad\times h_{\mathscr A}(y)h_{\mathscr A}(y)^\dagger h_{\Phi}(y)\nonumber\\
 &=h_{\Phi}(x)^\dagger U_\Gamma(x,y)h_{\Phi}(y),
 \label{eq:CHexact}
\end{align}
where $U_\Gamma=\mathcal P\exp(ig\int_\Gamma\mathscr A)$.
Thus $\mathcal C_{\rm rel}(x,y;\Gamma)$ defined in Eq.~\eqref{eq:Crel} is precisely equal to the \textbf{Higgs-frame Wilson-line correlator}:
\begin{equation}
 {C}_H(x,y;\Gamma)
 =\frac12\left\langle\operatorname{tr}\left[
 h_{\Phi}^\dagger(x)U_\Gamma(x,y)h_{\Phi}(y)
 \right]\right\rangle ,
 \label{eq:CHl}
\end{equation}
which is \emph{independent} of the choice of $h_{\mathscr A}$.
It cannot alone diagnose locking of an auxiliary confinement frame.

Moreover, if the fundamental scalar vanishes at an endpoint, the
normalized $h_\Phi=\hat\Theta$ is not defined. The \emph{unnormalized}
scalar--string observable remains regular there:
\begin{equation}
 C_{\Theta}(x,y;\Gamma)=\frac12\Big\langle
 \operatorname{tr}[\Theta^\dagger(x)U_\Gamma(x,y)
 \Theta(y)]\Big\rangle .
 \label{eq:regularscalarstringIII}
\end{equation}
On patches with nonzero amplitude it equals the normalized operator
weighted by $\rho(x)\rho(y)/2$ in the continuum convention
$\Theta=\rho\hat\Theta/\sqrt2$. Thus the normalized and
unnormalized correlators should not be equated when the radial mode
fluctuates. Line/endpoint/cusp subtractions must be fixed for either nonlocal string observable.

For an independently specified \emph{complete}, common-root
confinement-frame prescription $[C]$ we separately define the
untransported relative-frame correlator
\begin{align}
 q_C(x)&:=h_{\Phi}(x)^\dagger h_{\mathscr A}(x),\nonumber\\
 C_q^{[C]}(x,y)&:=\frac12\left\langle
 \operatorname{tr}[q_C(x)q_C(y)^\dagger]\right\rangle.
 \label{eq:frameonlyCq}
\end{align}
For a Wilson-tree prescription, Eq.~\eqref{eq:propositiontree}
shows that $C_q^{[T]}$ is exactly \textbf{the tree-path scalar--string correlator};
a change of tree is a change of its physical nonlocal operator,
not a gauge transformation.
For a genuine CDG partial color frame,
$C_q^{[C]}(x,y)$ is not lift-independent and must not be used without
extra structure.

Instead define,
\begin{align}
 r_C(x)&:=\hat\Theta(x)^\dagger \bm n_C(x)\hat\Theta(x),
 \label{eq:frameonlyr}\\
 C_{r,conn}(x,y)&:=C_r(x,y)-\frac12\operatorname{tr}
 [\langle r_C(x)\rangle\langle r_C(y)\rangle],
 \label{eq:frameonlyCr}
\end{align}
where $C_r(x,y)$ is defined
\begin{align}
 C_r(x,y)&:=\frac12\left\langle\operatorname{tr}
 [r_C(x)r_C(y)]\right\rangle,
 \label{eq:Crnonconn}
\end{align}
with the normalization
convention $\bm n_C(x)^2=\mathbf1$.
The subtraction in $C_{r,conn}(x,y)$ is made in the same state and with the same reduction-copy selection protocol used to compute $C_r(x,y)$.
The shorthand $C_{r,c}(x,y)$ used in later numerical discussion means precisely $C_{r,conn}(x,y)$ with this subtraction and the same copy protocol.

Long-distance alignment is tested with the nonconnected
$C_r(x,y)$ or the sign-even averaged norm
$\langle\|R_V\|_2^2\rangle$; the connected
$C_{r,\mathrm{conn}}(x,y)$ instead measures connected fluctuations
and their correlation length. In a symmetry-selected ordered pure
phase, the latter can decay to zero while the former tends to a
nonzero constant. A symmetric finite-volume ensemble and a chosen
copy/sign prescription must be distinguished from such a pure phase.

The subscript $conn$ in $C_{r,conn}(x,y)$ subtracts products of one-point
expectations in the chosen state. The connected two-point function
measures fluctuations and a connected correlation length; it is \emph{not}
by itself a long-range order parameter. In a symmetry-selected pure phase
with clustering, $C_{r,conn}(x,y)$ can decay even if a nonzero one-point
order parameter survives $\langle r_C\rangle$.

 In a finite symmetric ensemble $\langle r_C\rangle$ can vanish, in which case $C_r(x,y)$ and $C_{r,conn}(x,y)$ may coincide; this does not make their long-distance interpretations identical to those in a symmetry-selected pure phase.
A reproducible global-sign prescription, any custodial source, and the reduction copy algorithm are part of the definition.
The operator $r_C(x)$ is evaluated at a site but is generally nonlocal in the original links because $n_{\mathscr{A}}[U]$ is obtained by a global reduction.

Unlike a scalar string, $r_C(x)$ is generally nonlocal through the reduction functional, and its composite-operator continuum renormalization must be established separately.
Here $\bm n_{\mathscr{A}}(x)$ is a gauge-covariant direction obtained from an explicitly
specified reduction prescription. A Wilson-transport representation
$\bm n_{\mathscr{A}}(x)=h_{\mathscr{A}}(x)\sigma_3h_{\mathscr{A}}(x)^\dagger$ is allowed only with the covariant
root-axis/boundary convention of Eq.~\eqref{eq:rootcovaxis}.

For a CDG color direction $\bm n_{\mathscr{A}}(x)=h_{\mathscr{A}}(x)\sigma_3h_{\mathscr{A}}(x)^\dagger$, the lift is
not unique: $h_{\mathscr{A}}(x)\to h_{\mathscr{A}}(x)k_{C}(x)$ with
$k_{C}(x)\in U(1)_{\sigma_3}$. Therefore
\begin{equation}\label{eq:cdgstabilizerq}
 q_C(x)\longrightarrow  q_C(x)k_{C}(x),
\end{equation}
and a raw $\operatorname{tr}[q_C(x)q_C(y)^\dagger]$ generally
changes under independent $k_C(x),k_C(y)$: it is \emph{not}
an observable constructed from the CDG direction alone.
The local matrix $r_C(x)=q_C(x)\sigma_3q_C(x)^\dagger$ is independent
of this lift choice. The connected correlator $C_{r,conn}(x,y)$ defined above is hence the primary CDG relative-orientation two-point diagnostic.

A group-valued two-point alternative requires an explicitly specified
stabilizer transporter $u_H(x,y)$ satisfying
$u_H(x,y)\to k_C(x)^\dagger u_H(x,y) k_C(y)$ and the combination
$\operatorname{tr}[q_C(x)u_H(x,y)q_C(y)^\dagger]$; it is a
\emph{different} path-prescribed operator that requires a chosen
stabilizer-transport prescription, not merely the two endpoint
values of $\bm n_{\mathscr A}$.

After fixing line, endpoint, and cusp renormalization for a
specified path family, a normalized \textbf{scalar--string correlator} may exhibit a long-distance exponential falloff,
\begin{equation}
 C_H(R;\Gamma)\sim Z_H^{(\Gamma)}
 e^{-R/\xi_H^{(\Gamma)}} ,
 \label{eq:xirel}
\end{equation}
where $R$ and the path geometry are varied according to a fixed
prescription. The quantity $\xi_H^{(\Gamma)}$ is a
\emph{path-dependent} \textbf{scalar--string scale}, not automatically a
particle correlation length or an \textbf{intrinsic frame-locking length}.
For the Wilson tree $C_q^{[T]}$ and the corresponding $C_H$ are
identical operators and therefore cannot yield independent lengths.
A CDG-sensitive correlation length may instead be extracted from
$C_{r,conn}$, if its infinite-volume subtraction, spatial asymptotics,
copy selection, and renormalization are well defined.
No equality of these lengths, and no diagnosis of charged-sector confinement
or a thermodynamic phase, follows from their definitions.

\subsection{Gauss law and the physical meaning of complementarity}

In canonical notation, with $\mathscr E^j$ the non-Abelian electric field
and $\varrho_\Phi$ the matter color-charge density, the Gauss
constraint is
\begin{equation}
 \mathcal G(x):=(\mathscr D_j[\mathscr A]\mathscr E^j)(x)
             -\varrho_\Phi(x)\approx0 .
 \label{eq:gaussIII}
\end{equation}
Its smearing with test functions that vanish at the boundary
generates pure gauge redundancy and annihilates physical states.
Transformations with nontrivial boundary data instead require
boundary terms in their differentiable generators; the existence
and interpretation of the resulting charges depend on boundary
conditions and on the permitted physical symmetry group
\cite{GomesHopfmullerRiello,BorsboomPosthuma2026}.
A matter-frame dressing or a Wilson-line endpoint can furnish a
relational description of charged configurations, but a local
colored field is not thereby promoted to a local gauge-invariant
operator. In particular, the Gauss law alone is compatible with
screening, confinement, and Coulomb behavior: it \emph{cannot}
establish confinement without additional charged-sector information.

The confinement-oriented QRF framework \cite{KondoQRFConfinement2026}
and the Higgs-oriented construction \cite{KondoQRFBEH2026} can thus
be compared at fixed couplings without equating their nonlocal
observables. Where both descriptions are admissible, their
reparametrization of the \emph{same gauge-invariant operators}
preserves physical spectral poles and thresholds; it does not imply
equality of correlators of \emph{different} operators.
A Fradkin--Shenker interpolation is a change of the theory's
couplings, so the masses and overlaps in those channels can evolve
along the path, without a mandatory local thermodynamic singularity.
Nonlocal frame data may change differently, subject to the explicit
copy/path/renormalization tests above. The regimes are therefore
not asserted to have identical spectra at different couplings.

\section{Fundamental scalar in the continuum}
\label{sec:fundcont}

\subsection{Complete Higgs frame and fixed radial length}

For Hermitian generators and Euclidean signature, take
\begin{align}
 &S=S_{\rm YM}[\mathscr A]+\int d^4x\,
 \bigl[(D_\mu\Phi)^\dagger D_\mu\Phi
        +V(\Phi^\dagger\Phi)\bigr], \nonumber\\
 &S_{\rm YM}[\mathscr A]
 =\frac12\int d^4x\,
  \operatorname{tr}\!\left(\mathscr F_{\mu\nu}[\mathscr A]
                    \mathscr F_{\mu\nu}[\mathscr A]\right).
 \label{eq:Sfund}
\end{align}
For an $SU(2)$ fundamental doublet $\Phi(x)$, define the $2\times2$ matrix-valued scalar field
\begin{equation}
 \Theta(x)=(i\sigma_2\Phi(x)^{*},\,\Phi(x)).
 \label{eq:PhiMatrix}
\end{equation}
On a patch where $\Phi^{\dagger}\cdot\Phi\neq0$ one may write the polar decomposition
\begin{align}
 &\Theta(x)=\frac{\rho(x)}{\sqrt{2}}\hat{\Theta}(x),
\nonumber\\
 &\rho(x):=\sqrt{2\Phi(x)^{\dagger}\cdot\Phi(x)} \ge 0,
 \quad
 \hat{\Theta}(x)\in SU(2),
 \label{eq:polar}
\end{align}
where  $\hat\Theta$ satisfies
\begin{equation}
 \hat\Theta^\dagger\hat\Theta=\bm1,
 \qquad \det\hat\Theta=1.
 \label{eq:apppolar}
\end{equation}
Under the left local $SU(2)$ action,
\begin{equation}
\Theta(x)\mapsto\Omega(x)\Theta(x), \quad \Omega(x)\in SU(2),
\end{equation}
therefore $\hat\Theta(x)$ is a complete local Higgs frame only
where $\rho(x)>0$.  The normalized matrix $\hat\Theta$ is undefined at scalar
zeros $\rho(x)=0$, whereas $\Theta$ itself is regular.

For a potential $V$ depending only on $\Phi^\dagger\Phi$, the theory also has a \emph{global} right custodial action $\hat\Theta\mapsto\hat\Theta R$ ($R\in SU(2)$), distinct from the local left gauge redundancy.
Consequently $\widetilde{\mathscr W}_\mu$ transforms in the adjoint of the custodial group; its components are not custodial singlets.

With $\rho=\sqrt{2\Phi^\dagger\Phi}$, a fixed-length
constraint $\Phi^\dagger\Phi=v^2/2$ is precisely $\rho=v$;
only $\hat\Theta$ then fluctuates among the scalar variables.
Here $v$ specifies a fixed radius, not a gauge-variant vacuum
expectation value.  When the radial mode fluctuates, the same
symbol denotes a chosen expansion background, and its physical
interpretation requires gauge-invariant observables.

When the matter frame is chosen
\begin{equation}
h(x)=\hat{\Theta}(x),
\end{equation}
the relational vector field is obtained as
\begin{align}
 \mathscr{A}^{\hat{\Theta}}_{\mu}(x)
 :=&\hat{\Theta}(x)^{\dagger}\mathscr{A}_{\mu}(x)\hat{\Theta}(x)
 +\frac{i}{g}\hat{\Theta}(x)^{\dagger}\partial_{\mu}\hat{\Theta}(x) \nonumber\\
 =&\frac{i}{g}\hat{\Theta}(x)^{\dagger}D_{\mu}[\mathscr{A}]\hat{\Theta}(x):=\widetilde{\mathscr{W}}_{\mu}(x).
 \label{eq:WtildeCont}
\end{align}
It is gauge invariant with respect to the local left action and is exactly the gauge field expressed in the scalar QRF.

\textbf{The fixed-length scalar kinetic term becomes a positive quadratic form in $\widetilde{\mathscr{W}}_{\mu}$, reproducing the standard BEH vector mass without using a gauge-variant scalar expectation value as a physical order parameter} \cite{Kondo2016,Kondo2017,Kondo2018,KondoQRFBEH2026}.

Using
\begin{equation}
 D_\mu[\mathscr A]\Theta
 =\frac{1}{\sqrt2}(\partial_\mu\rho)\hat\Theta
 +\frac{\rho}{\sqrt2}D_\mu[\mathscr A]\hat\Theta.
 \label{eq:DPhi}
\end{equation}
and
$\hat\Theta^\dagger D_\mu[\mathscr{A}]\hat\Theta=-ig\widetilde{\mathscr W}_\mu$
and the anti-Hermiticity of $\hat\Theta^\dagger\partial_\mu\hat\Theta$,
the mixed radial--angular term drops out after taking the trace.
Hence we find
\begin{equation}
 \frac12\operatorname{tr}[(D_\mu[\mathscr{A}]\Theta)^\dagger(D_\mu[\mathscr{A}]\Theta)]
 =\frac12(\partial_\mu\rho)^2
 +\frac14g^2\rho^2
  \operatorname{tr}(\widetilde{\mathscr W}_\mu\widetilde{\mathscr W}_\mu).
 \label{eq:kinrewrite}
\end{equation}
This normalization agrees with the doublet identity
$\frac12\operatorname{tr}[(D_\mu[\mathscr{A}]\Theta)^\dagger D_\mu[\mathscr{A}]\Theta]
=(D_\mu[\mathscr{A}]\Phi)^\dagger \cdot D_\mu[\mathscr{A}]\Phi$.
With the doublet convention $\Phi^\dagger\cdot\Phi=\frac{\rho^2}{2}$,
the corresponding tree-level vector stiffness is
\begin{equation}
   (D_\mu[\mathscr{A}]\Phi)^\dagger \cdot D_\mu[\mathscr{A}]\Phi
 =\frac12 (\partial_\mu\rho)^2+\frac{g^2}{8}\rho^2
 \widetilde{\mathscr W}_\mu^A\widetilde{\mathscr W}_\mu^A.
 \label{eq:treevectorstiffnessadd}
\end{equation}
For $\rho=v$, comparison with
$\frac12m_{W,0}^2\widetilde{\mathscr W}_\mu^A
\widetilde{\mathscr W}_\mu^A$ gives the tree-level mass $m_{W,0}=gv/2$.
This is a quadratic bare stiffness, not an exact assertion of a
narrow physical vector pole at arbitrary coupling. The latter
must be extracted from a renormalized, gauge-invariant vector
interpolating operator and its spectral representation.
The physical mass is identified through the relevant invariant
correlator's pole or threshold (in a suitable channel).

The relational field strength is defined by
\begin{equation}
 \mathscr{F}_{\mu\nu}[\widetilde{\mathscr{W}}](x)
 =\hat{\Theta}(x)^{\dagger}\mathscr{F}_{\mu\nu}[\mathscr{A}](x)\hat{\Theta}(x) .
 \label{eq:Ftransform}
\end{equation}
The Yang--Mills term transforms just as directly under the same field-dependent frame change:
\begin{equation}
 \mathscr F_{\mu\nu}[\widetilde{\mathscr W}]
 =\hat\Theta^\dagger\mathscr F_{\mu\nu}[\mathscr A]\hat\Theta,
 \qquad
 \operatorname{tr}\mathscr F_{\mu\nu}[\mathscr A]^2
 =\operatorname{tr}\mathscr F_{\mu\nu}[\widetilde{\mathscr W}]^2.
 \label{eq:FtransformAppendix}
\end{equation}
Combining Eqs.~\eqref{eq:kinrewrite} and \eqref{eq:FtransformAppendix} gives the exact patchwise rewriting
\begin{align}
 S=&\int d^4x\left[
 \frac12(\partial_\mu\rho)^2
 +\frac{g^2}{4}\rho^2
 \operatorname{tr}(\widetilde{\mathscr W}_\mu\widetilde{\mathscr W}_\mu)
 +V\!\left(\frac{\rho^2}{2}\right)
 \right]\nonumber\\
 &+S_{\rm YM}[\widetilde{\mathscr W}],\quad
 S_{\rm YM}[\widetilde{\mathscr W}]
 =\frac12\int d^4x\,
 \operatorname{tr}\mathscr F_{\mu\nu}[\widetilde{\mathscr W}]^2.
 \label{eq:exactradialrewrite}
\end{align}
The positive Yang--Mills sign is the one compatible with a stable Euclidean weight $e^{-S}$.  The rewriting is exact on every patch with $\rho>0$; at $\rho=0$ the original Cartesian scalar variables, or an explicitly patchwise description, must be used.

With a gauge-derived frame $h_{\mathscr A}$, define
\begin{equation}
 \mathscr{A}^{h_{\mathscr A}}_{\mu}(x):=h_{\mathscr A}(x)^{\dagger}\mathscr{A}_{\mu}(x)h_{\mathscr A}(x)
 +\frac{i}{g}h_{\mathscr A}(x)^{\dagger}\partial_{\mu}h_{\mathscr A}(x).
 \label{eq:qfund}
\end{equation}
By introducing
\begin{equation}
\boxed{ q(x)=\hat{\Theta}(x)^{\dagger} h_{\mathscr A}(x)},
 \label{eq:qfund-higgs-orientation}
\end{equation}
we obtain the relationship between $\mathscr{A}^{\hat{\Theta}}_{\mu}(x)=\widetilde{\mathscr{W}}_{\mu}(x)$ and $\mathscr{A}^{h_{\mathscr A}}_{\mu}(x)$:
\begin{equation}
 \widetilde{\mathscr{W}}_{\mu}(x)
 =q(x)\mathscr{A}^{h_{\mathscr A}}_{\mu}(x)q(x)^{\dagger}
 +\frac{i}{g}q(x)\partial_{\mu}q(x)^{\dagger}.
 \label{eq:qWidentity}
\end{equation}
This is the continuum form of the exact lattice identity that will underlie the interpretation of Ref.~\cite{Ikeda2024}.

A polynomial, gauge-invariant vector interpolating field
(with custodial index $A$) is available even at scalar zeros:
\begin{align}
 \mathcal J_\mu^A(x)
 :=\frac{i}{g}\operatorname{tr}\!\left[
 \sigma^A\Theta^\dagger D_\mu[\mathscr A]\Theta\right]
 =\frac{\rho^2}{2}\,\widetilde{\mathscr W}_\mu^A
 \qquad (\rho>0).
 \label{eq:regularcustodialcurrent}
\end{align}
The same decomposition also makes the regular custodial-vector operator transparent.  One has
\begin{equation}
 \Theta^\dagger D_\mu[\mathscr A]\Theta
 =\frac{1}{2}\rho\partial_\mu\rho \bm1
  -\frac{1}{2}ig\rho^2\widetilde{\mathscr W}_\mu.
 \label{eq:appRegularCurrentDerivation}
\end{equation}
Multiplication by $\sigma^A$, followed by a trace and by $i/g$, removes the radial identity term and gives
$\mathcal J_\mu^A=\rho^2\widetilde{\mathscr W}_\mu^A/2$.
Thus $\mathcal J_\mu^A$ remains a polynomial in the original scalar and gauge variables even when the normalized frame itself ceases to exist.  This algebraic identity does not by itself imply equality of physical poles; a spectral-overlap argument is still required.
The trace is real because its identity-component contribution
vanishes.  Although the normalized field
$\widetilde{\mathscr W}_\mu$ can fail at zeros,
$\mathcal J_\mu^A$ remains a regular polynomial in the original
scalar and gauge fields.  It is the corresponding custodial-triplet
channel, rather than a bare mass coefficient or an arbitrarily
defined frame correlator, whose spectral function must be analyzed.
This is consistent with the composite-operator viewpoint of
Refs.~\cite{FMS1980,FMS1981}.

This is also the natural point of contact with the FMS construction. FMS starts from a gauge-invariant composite and, in a suitable Higgs regime, relates its pole structure to a gauge-fixed expansion. The present QRF organization instead keeps the dynamical matter frame explicit and compares it with an independently constructed gauge-field frame. The physical pole is frame independent, whereas a line-dressed relative-frame correlator may carry path-dependent perimeter contributions and must not be identified with that pole without an operator-overlap and renormalization analysis.

\subsection{Restoring the radial mode: exact rewriting}
\label{subsec:radialfund}

We now complete the radial measure and regular-bilinear
dictionary of Sec.~III\,A. The polar kinetic decomposition and the
stiffness/current identities need not be derived a second time;
the remainder of this subsection records their consequences for
the exact measure and nonlocal observables.

The measure follows from the same polar geometry rather than being an independent assumption.  For $\Phi=(\phi_1,\phi_2)^T\in\mathbb C^2$, the radius in $\mathbb R^4$ is $\rho/\sqrt2$, while the angular manifold is $S^3\simeq SU(2)$.  With normalized Haar measure $d\mu_H$,
\begin{equation}
 d\Omega_3=2\pi^2\,d\mu_H(\hat\Theta).
 \label{eq:polarmeasureapp}
\end{equation}
Consequently,
\begin{equation}
 d^4\Phi
 =(\rho/\sqrt2)^3d(\rho/\sqrt2)d\Omega_3
 =\frac{\pi^2}{2}\rho^3d\rho\,d\mu_H(\hat\Theta).
 \label{eq:measurepolar}
\end{equation}
The overall constant cancels in normalized expectation values.  On a finite lattice the accompanying field-dependent rotation of each gauge link has unit Haar-measure Jacobian; integration over $\hat\Theta$ supplies the gauge volume in gauge-invariant expectation values wherever the chart is nonsingular.  The factor $\rho^3$ must not be discarded.  In the continuum, regulator dependence, patch boundaries, operator counterterms, contact terms, and the treatment of zeros must still be specified consistently.

This derivation also clarifies the interpretation of the quadratic term.  The coefficient $g^2\rho(x)^2/8$ is a local vector \emph{stiffness}; it is not a position-dependent observable pole mass.  Likewise, a radial curvature extracted from $V$ is a bare small-fluctuation scale until a gauge-invariant spectral channel is analyzed.

At $\rho=0$ the normalized frame $\hat\Theta$ is undefined, and therefore the relative frame $q=\hat\Theta^\dagger h_{\mathscr A}$ fails as well. This is a failure of the normalized QRF, not a singularity of the original Cartesian scalar.  A regular variable is instead the unnormalized relational scalar
\begin{equation}
 \Psi(x)=\Theta(x)^\dagger h_{\mathscr A}(x)
 =\frac{\rho(x)}{\sqrt2}\,q(x).
 \label{eq:Psi}
\end{equation}
The factor $1/\sqrt2$ follows from the continuum polar convention in Eq.~\eqref{eq:polar}; the independently normalized lattice amplitude in Eq.~\eqref{eq:lattpolar} instead has $\Psi_x=\rho_xq_x$.  Its two-point function is
\begin{align}
 \mathcal C_\Psi(x,y)
 &=\frac12\left\langle
 \operatorname{tr}[\Psi^\dagger(x)\Psi(y)]\right\rangle\nonumber\\
 &=\frac14\left\langle\rho(x)\rho(y)
 \operatorname{tr}[q(x)q(y)^\dagger]\right\rangle,
 \label{eq:CPsi}
\end{align}
and remains finite when configurations contain scalar zeros.  In this sense $\rho$ measures the local reliability or weight of the normalized Higgs frame: a region of small $\rho$ is a region in which that frame is a poor reference system.

Finally, the same expression shows why a general gauge-derived frame must not be silently replaced by a Wilson-tree frame.  For any complete $h_{\mathscr A}$,
\begin{equation}
 \mathcal C_\Psi(x,y)=\frac12\left\langle
 \operatorname{tr}\!\left\{\Theta^\dagger(x)
 h_{\mathscr A}(x)h_{\mathscr A}(y)^\dagger\Theta(y)\right\}
 \right\rangle.
 \label{eq:CPsiBilocal}
\end{equation}
If $h_{\mathscr A}(x)=U(P_x)$ is a common-root Wilson frame, the middle factor reduces to the prescribed transporter
$U(P_x)U(P_y)^\dagger=U(P_{y\to x})$, and $\mathcal C_\Psi$ is precisely that scalar--string correlator.  For an independently chosen CDG/reduction frame, the middle factor need not be a Wilson transporter on a single prescribed path.  If only a CDG color direction is supplied, its lift also retains a local $U(1)$ ambiguity.  Hence the raw group-valued bilocal must not be identified with the stabilizer-invariant CDG relative-color correlator of Sec.~\ref{sec:continuumgeneral}.  In the Wilson-rooted case the trace is nevertheless invariant under the common root rotation, even though $\Psi$ itself transforms at the root.

\subsection{Higgs-like side: bare stiffness and radial curvature scales}

Consider a smooth patch with $\rho(x)=v+H(x)$, $|H|\ll v$, with $H$ canonically normalized.  The exact rewriting in Eq.~\eqref{eq:exactradialrewrite} immediately separates two different small-fluctuation scales.  The quadratic vector term gives the tree-level stiffness
\begin{equation}
 M_{\mathscr W}\sim \frac12gv,
 \label{eq:MW}
\end{equation}
whereas the radial scale follows from the curvature of the same potential that appears in the full action.  Writing $s=\rho^2/2$, one has
\begin{equation}
 \frac{dV(s)}{d\rho}=\rho V'(s),\qquad
 \frac{d^2V(s)}{d\rho^2}=V'(s)+\rho^2V''(s).
 \label{eq:appcorrectradmass}
\end{equation}
Therefore, at a stationary background $V'(v^2/2)=0$,
\begin{equation}
 m_{H,0}^2
 =\left.\frac{d^2V(\rho^2/2)}{d\rho^2}\right|_{\rho=v}
 =v^2V''(v^2/2).
 \label{eq:MH}
\end{equation}
Away from a stationary background the curvature is instead
$V'(v^2/2)+v^2V''(v^2/2)$.  Thus $m_{H,0}$ is the canonically normalized radial curvature mass, not automatically a nonperturbative scalar pole mass, and $M_{\mathscr W}$ is likewise a bare stiffness rather than an observable vector pole.  A physical mass requires a gauge-invariant interpolating operator and its spectral representation.  The regular correlator $\mathcal C_\Psi$ can therefore contain more than one scale, which is why a single ``Higgs correlation length'' is insufficient once the radial mode fluctuates.

The same curvature formula also identifies the fixed-length limit.  At fixed chosen radius $v$, the radial curvature must be taken appropriately large.  Since $\rho^2=2\Phi^\dagger\Phi$, the conventional doublet normalization
$\langle\Phi^\dagger\Phi\rangle=v_{\rm SM}^2/2$
corresponds to $\rho=v_{\rm SM}$ in the present convention and gives the bare vector scale $m_{W,0}=gv_{\rm SM}/2$.  The fixed-length limit and the large-vector-stiffness limit are distinct operations and need not commute with the continuum or infinite-volume limits.

The relation between frame locking and the vector stiffness is encoded directly in the exact change-of-frame identity. Rearranging Eq.~\eqref{eq:qWidentity} for $q=h_\Phi^\dagger h_{\mathscr A}$ gives
\begin{equation}
 \partial_\mu q(x)
 =ig\,\widetilde{\mathscr W}_\mu(x)q(x)
 -ig\,q(x)\mathscr A_\mu^{h_{\mathscr A}}(x).
 \label{eq:dq}
\end{equation}

Let $z(s)$, $0\le s\le1$, parametrize a path $\Gamma:y\to x$, and set
\begin{align}
 \mathcal U_{\mathscr A}(x,y;\Gamma)
 &=\mathcal P\exp\!\left(ig\int_y^x dz^\mu\,
   \mathscr A_\mu^{h_{\mathscr A}}(z)\right),
 \label{eq:UCapp}\\
 \mathcal U_{\mathscr W}(x,y;\Gamma)
 &=\mathcal P\exp\!\left(ig\int_y^x dz^\mu\,
   \widetilde{\mathscr W}_\mu(z)\right).
 \label{eq:UW}
\end{align}

Along the path, Eq.~\eqref{eq:dq} becomes
$\dot q=ig\dot z^\mu\widetilde{\mathscr W}_\mu q
      -igq\dot z^\mu\mathscr A_\mu^{h_{\mathscr A}}$.
The product
$\mathcal U_{\mathscr W}(z(s),y)q(y)
 \mathcal U_{\mathscr A}(z(s),y)^\dagger$
obeys this first-order equation with initial value $q(y)$. Uniqueness of the path-ordered solution therefore gives
\begin{equation}
 q(x)=\mathcal U_{\mathscr W}(x,y;\Gamma)
 q(y)\mathcal U_{\mathscr A}(x,y;\Gamma)^\dagger.
 \label{eq:qtransport}
\end{equation}
Thus the right factor is the inverse \emph{confinement-frame} holonomy $\mathcal U_{\mathscr A}^\dagger$, not the inverse Higgs-frame holonomy.

For a Wilson frame defined on a prescribed rooted tree,
$\mathscr A^{h_{\mathscr A}}$ has vanishing parallel component only on the chosen tree edges.  Hence $\mathcal U_{\mathscr A}=\bm1$ for a path lying entirely in that tree, and Eq.~\eqref{eq:qtransport} reduces there to a relation controlled by the $\widetilde{\mathscr W}$ holonomy.  This is a property of that prescription, not a statement for an arbitrary curve or a closed Wilson loop; non-tree cycles still detect curvature.

The large-stiffness expansion is therefore an expansion of a \emph{renormalized, fixed-path} relational holonomy, not a direct expansion of a local particle pole.  Increasing $M_{\mathscr W}$ can suppress connected line fluctuations under additional scaling assumptions, but $M_{\mathscr W}\to\infty$ alone does not imply the operator identity
$\mathcal U_{\mathscr W}\to\bm1$.

The order of limits is consequently essential.  The radial-curvature limit producing a fixed-length scalar theory is distinct from the large-vector-stiffness limit.  Moreover, the limits
$M_{\mathscr W}\to\infty$, path length $L\to\infty$, volume $V\to\infty$, and lattice spacing $a\to0$
need not commute.  A long Wilson/tree segment can retain a perimeter contribution proportional to $L$ even when local vector correlations are short-ranged.  A genuine long-distance frame-locking statement therefore requires fixed-path renormalization and explicit control of the large-stiffness, long-distance, infinite-volume, and continuum limits rather than an inference from the bare stiffness alone.

\subsection{Fradkin--Shenker continuity in fundamental QRF language}

Away from zeros a fundamental scalar supplies a complete
frame with no residual \emph{local} stabilizer. Dynamical
fundamental matter explicitly breaks the electric center one-form symmetry \cite{GKSW2015}: fundamental Wilson lines can end on the matter.
Neither observation is itself an analyticity theorem.
Fradkin--Shenker continuity establishes an analytic connected
region for the specified local gauge-invariant observables in
fixed-length lattice gauge--Higgs models
\cite{FradkinShenker,OsterwalderSeiler}; it does not prohibit
all possible transition lines elsewhere in parameter space,
and does not automatically extend unchanged to every continuum
radial potential or nonlocal dressing.

The relative-frame formulation is compatible with that
continuity, but makes a separate, testable proposal: specified
Wilson-rooted or CDG frame diagnostics may exhibit different
infrared behavior as couplings change. Such a change can be
smooth, or can be nonanalytic for a suitably defined nonlocal
observable even where the local observables covered by the theorem
are analytic. No latter singularity follows from the frame
identity alone. A normalized scalar frame can fail at zeros;
small radial weights do not by themselves count topological
defects.  The observables, paths, copy prescription, and line
renormalization must therefore be fixed before identifying a
``locking'' regime. The physically separate test is whether
its infrared behavior correlates with a charged-sector isolation
energy or gauge-invariant spectral data
\cite{KondoQRFConfinement2026}.

\section{Lattice fundamental scalar  formulation}
\label{sec:lattice}

\subsection{Fundamental scalar with a fluctuating radius}

On an oriented hypercubic lattice the link variable $U_{x,\mu}\in SU(2)$ transforms under the gauge transformation as
\begin{equation}
U_{x,\mu}\mapsto\Omega_xU_{x,\mu}\Omega_{x+\hat\mu}^\dagger .
\end{equation}
The Wilson gauge action is written in terms of the gauge-invariant plaquette variable $U_P$ formed from the link variables $U_{x,\mu}$ as
\begin{equation}\label{eq:gaugeaction}
 S_{\rm G}[U]=\frac{\beta}{2}\sum_P\operatorname{Re}\operatorname{tr}(\mathbf1-U_P),\qquad \beta=\frac4{g^2}\quad(D=4).
\end{equation}
In fact, with links $U_{x,\mu}=\exp(iga\mathscr A_\mu+O(a^2))$
and generators $T^A=\sigma^A/2$, the standard plaquette matching
for a $D$-dimensional hypercubic lattice gives
\begin{equation}\label{eq:betaD}
 \beta_D=\frac{4a^{D-4}}{g_D^2},\qquad
 [g_D^2]=\mathrm{mass}^{4-D}.
\end{equation}
Hence $\beta_4=4/g_4^2$.

Consider a matrix-valued fundamental scalar $\Theta$ written as
\begin{equation}
 \Theta_x=\rho_x\hat{\Theta}_x,
 \qquad \rho_x\ge0,
 \qquad \hat{\Theta}_x\in SU(2).
 \label{eq:lattpolar}
\end{equation}
The total action of $SU(2)$ gauge-Higgs model with the fundamental scalar field is given by
\begin{align}
 &S=S_{\rm G}[U]+S_H[U,\rho,\Theta]+\sum_x V(\rho_x),\nonumber\\
 &S_H[U,\rho,\hat{\Theta}]=-\frac{\kappa}{2}\sum_{x,\mu}
 \rho_x\rho_{x+\hat\mu}\,
 \operatorname{Re}\operatorname{tr}
 \left(\hat{\Theta}_x^{\dagger}U_{x,\mu}\hat{\Theta}_{x+\hat\mu}\right).
 \label{eq:latactionrad}
\end{align}
Define the scalar-frame link variable
\begin{equation}
 \widetilde W_{x,\mu}
 =\hat{\Theta}_x^{\dagger}U_{x,\mu}\hat{\Theta}_{x+\hat\mu}.
 \label{eq:Wlattice}
\end{equation}
It is invariant under the local left gauge transformation.  At fixed radial length this is exactly the gauge-invariant link variable used in the lattice gauge-independent BEH construction and in Appendix A of Ref.~\cite{Ikeda2024}.
Then the scalar hopping term $S_H$ becomes simply a function of $\rho_x\rho_{x+\hat\mu}$ and $\widetilde W_{x,\mu}$:
\begin{align}\label{eq:SH}
S_H[\widetilde W,\rho]=-\frac\kappa2\sum_{x,\mu}\rho_x\rho_{x+\hat\mu}
 \operatorname{Re}\operatorname{tr}(\widetilde W_{x,\mu}) .
\end{align}
This sign is consistent with the positive hopping contribution in $e^{-S}$ used below. For a fluctuating radial mode, rewriting the hopping term with $\operatorname{tr}(\mathbf1-\widetilde W)$ does not differ merely by an additive constant because $\rho_x\rho_{x+\hat\mu}$ is dynamical; the difference is an additive constant only at fixed radial length.

Let $h_{\mathscr{A},x}[U]$ be a gauge-derived frame and define the relational link variable $U^{\mathscr{A}}_{x,\mu}$ by
\begin{equation}
 U^{\mathscr{A}}_{x,\mu}
 =h_{\mathscr{A},x}^{\dagger}U_{x,\mu}h_{\mathscr{A},x+\hat\mu}.
 \qquad
 \label{eq:UCqx}
\end{equation}
Then, configuration by configuration,
we find the relationship between two relational gauge fields relative to different frames:
\begin{equation}
 \widetilde W_{x,\mu}
 =q_xU^{\mathscr{A}}_{x,\mu}q_{x+\hat\mu}^{\dagger}, \quad
 q_x=\hat\Theta_x^{\dagger}h_{\mathscr{A},x}.
 \label{eq:exactlinkidentity}
\end{equation}
This is the lattice counterpart of Eq.~\eqref{eq:qWidentity}.  It is the central exact identity linking the confinement and Higgs QRFs.

For fluctuating $\rho_x$, the regular relative scalar is written as
\begin{equation}
 \Psi_x=\rho_x q_x.
 \label{eq:PsiLattice}
\end{equation}
The normalized $q_x$ is unreliable at small $\rho_x$, while $\Psi_x$ remains regular.

For an unfrozen scalar the radial mode is an explicit dynamical, gauge-invariant variable with a separate mass and the probability distribution of small $\rho$ controls the stability of the normalized Higgs frame.
A useful scalar-zero diagnostic is specified by
\begin{equation}
 P_{\epsilon}=\operatorname{Prob}(\rho_x<\epsilon),
 \label{eq:Peps}
\end{equation}
which can be measured together with conditional $q$ correlators and the unconditional $\Psi$ correlator.

The flat four-dimensional site measure for the scalar matrix is
\begin{equation}\label{eq:measure}
 d^4\Phi_x=\mathcal N\rho_x^3d\rho_x\,d\mu(\hat\Theta_x),\quad \rho_x\ge0, \quad \hat\Theta_x \in SU(2),
\end{equation}
with normalized Haar measure $d\mu(\hat\Theta)$ on $SU(2)$. Haar invariance maps all $U_{x,\mu}$ to $\tilde W_{x,\mu}$ without changing the link Jacobian, and the plaquette action depends only on the plaquette variable $W_P$ constructed from the link variables $\widetilde W_{x,\mu}$. After integrating the pure gauge orientations, the scalar-frame partition function is, up to the gauge-volume normalization, written as
\begin{align}\label{eq:Zscalar}
 &Z= \int\!D\widetilde W\!\prod_x\rho_x^3d\rho_x\,e^{-S_{\rm G}[\widetilde W]-\sum_xV(\rho_x)}\nonumber\\
 &\quad\times\exp\left[\frac\kappa2\sum_{x,\mu}\rho_x\rho_{x+\hat\mu}
 \operatorname{Re}\operatorname{tr}(\widetilde W_{x,\mu}) \right], \nonumber\\
  &S_{\rm G}[\widetilde W]= \frac{\beta}{2}\sum_P\operatorname{Re}\operatorname{tr}(\mathbf1-\widetilde W_P) .
\end{align}

A \textbf{Wilson-tree frame}, $h_{{\mathscr A},x}$, formed by parallel transport from a fixed root, gives $U^{\mathscr A}_{x,\mu}=\mathbf1$ on a tree edge. Then
\begin{equation}\label{eq:treeid}
 \widetilde W_{x,\mu}=q_x q_{x+\hat\mu}^\dagger\quad\text{on tree edges},
\end{equation}
where the common root transformation cancels in two-point traces. The gauge-invariant two-point diagnostics are
\begin{align}\label{eq:CqCPsi}
 C_q^T(x,y)&=\frac12\langle\operatorname{tr}[q_x q_y^\dagger]\rangle,\\
 C_\Psi^T(x,y)&=\frac12\langle\rho_x\rho_y\operatorname{tr}[q_x q_y^\dagger]\rangle.
\end{align}
A path- or tree-dependent correlator is a legitimate observable after its prescription is specified, but it must not automatically be called a universal local order parameter. Small $\kappa$ probes gauge links in a hopping expansion; large $\kappa$ probes a Higgs-frame link near unity. The exact $\beta=0$ bridge between the two is given in
Sec.~\ref{subsec:beta0}.

\subsection{Fixed-length fundamental model}
\label{subsec:IkedaVariable}

In the radially fixed model set $\rho_x=v$ and absorb $v$ into the hopping parameter $\kappa$: $\gamma:=\kappa v^2$.  Suppose the gauge-field QRF $h_{{\mathscr A},x}$ determines a color direction
\begin{equation}
 \bm n_{\mathscr{A},x}=h_{{\mathscr A},x}\sigma_3h_{{\mathscr A},x}^{\dagger},
 \label{eq:nfromh}
\end{equation}
For a CDG frame this equation is a local lift of an independently
fixed covariant direction, unique only modulo its $U(1)$ stabilizer.
For a Wilson-root frame the fixed $\sigma_3$ requires the root
gauge convention of Eq.~\eqref{eq:rootcovaxis}, or replacement by
a covariantly transforming root axis $n_0$.
These two constructions must not be identified without specifying
these extra data.

The gauge-invariant composite operator evaluated at a site and used in
Ref.~\cite{Ikeda2024} is, in matrix notation,
\begin{equation}
 \boxed{ r_x=\hat\Theta_x^{\dagger}\bm n_{\mathscr{A},x}\hat\Theta_x } .
 \label{eq:rxIkeda}
\end{equation}
Thus \textbf{$r_x$ is the gauge-field-derived color direction $\bm n_{\mathscr{A},x}$ read in the scalar frame $h_x=\hat\Theta_x$.}  Its spatial average
\begin{equation}
 \boxed{ R=\frac{1}{N_{\rm site}}\sum_x r_x, \quad R^A=\frac12\operatorname{tr}(\sigma^AR) }.
 \label{eq:R}
\end{equation}

\paragraph*{Two norms in the lattice fundamental example.}
To retain both observables used in the notation of
Ref.~\cite{KondoQRFConfinement2026}, write
$R=\sum_{A=1}^{3}R^{A}\sigma_A$ and define
\begin{align}
 \|R\|_1&:=\sum_{A=1}^{3}|R^{A}|,
 \label{eq:ep280-Rnorm1}\\
 \|R\|_2&:=
 \left[\frac12\operatorname{tr}(R^\dagger R)\right]^{1/2}
 =\left[\sum_{A=1}^{3}(R^{A})^2\right]^{1/2}.
 \label{eq:ep280-Rnorm2}
\end{align}
Both quantify the magnitude of the \emph{volume-averaged}
relative color direction. They obey
$\|R\|_2\leq\|R\|_1\leq\sqrt{3}\,\|R\|_2$,
so their vanishing in the thermodynamic limit is equivalent.
There is, however, a mathematical qualification: $\ell^2$ norm $\|R\|_2$
is strictly invariant under the global right $SU(2)$ rotation,
whereas the componentwise $\ell^1$ norm $\|R\|_1$ depends on
which custodial coordinate axes are used. Hence the latter
must not be called an exactly rotation-invariant norm. Its
finite-volume-subtracted expectation can nevertheless be
studied with the explicitly specified coordinate and symmetry
averaging prescription of Ref.~\cite{Ikeda2024}.
The observables reported there include subtracted magnitudes and their
susceptibilities, not the expectation of a locally color-charged scalar,
and rotationally insensitive norms $\| R \|$ of $R$ measure persistent relative alignment.
The right/global scalar symmetry acts on this relational orientation but \textbf{the left local gauge redundancy has already canceled}.  This is why a nonzero magnitude of $R$ can be compatible with Elitzur's theorem.

Although $r_x$ is assigned to a single lattice site,
its reduction-derived constituent $n_x[U]$ normally depends on
links throughout the volume. Accordingly $r_x$ need not be a
strictly bounded-support local observable of the original
lattice action, and local OSFS analyticity does not automatically
apply to its thermodynamic limit.

For $r_x=r_x^A\sigma_A$, $r_x^2=\mathbf1$, and
$R_V=V^{-1}\sum_{x\in V}r_x$, the matrix identity
$\frac12\operatorname{tr}(R_V^2)
 =V^{-2}\sum_{x,y}\frac12\operatorname{tr}(r_xr_y)$
holds configuration by configuration. After ensemble averaging with
one fixed, gauge-equivariant reduction-copy prescription $\mathcal S$,
\begin{align}\label{eq:R-Cn-sum-A2}
 \big\langle\|R_V\|_2^2\big\rangle_{\mathcal S}
 =\frac{1}{V^2}\sum_{x,y\in V}
 \left\langle\frac12\operatorname{tr}(r_xr_y)
 \right\rangle_{\mathcal S}
 =\frac{1}{V}\sum_{z\in V}C_n^{[\mathcal S]}(z),
\end{align}
where $C_n^{[\mathcal S]}(z):=V^{-1}\sum_x
\langle\tfrac12\operatorname{tr}(r_xr_{x+z})\rangle_{\mathcal S}$;
for a translation-invariant ensemble and copy rule this equals the
correlator at any reference site. If $C_n(z)$ clusters to $m^2>0$
and the Ces\`aro spatial mean converges, the squared norm tends to
$m^2$. If instead correlations are summable uniformly in volume,
it is $O(V^{-1})$. The observable in IKKS is a \emph{subtracted
magnitude} such as $\langle\|R\|_n\rangle_{\rm sub}$,
not the squared norm in this exact identity; finite-volume
subtractions must be performed separately. The equation gives a
sharp same-ensemble numerical consistency test, not equality of
the two differently normalized observables.
We can also show that $\|r_x\|_2=1$ implies $0\le\|R_V\|_2\le1$ and hence
\begin{equation}
\left\langle\|R_V\|_2^2\right\rangle
\le\left\langle\|R_V\|_2\right\rangle
\le\sqrt{\left\langle\|R_V\|_2^2\right\rangle}.
\end{equation}
Therefore, the vanishing of the raw $L^2$ norm is equivalent to the vanishing of the mean square.

The normalized matrix called $\Theta$ in
Ref.~\cite{KondoQRFConfinement2026} is the $\widehat\Theta$
of the present paper:
\begin{equation}
 \frac{\bigl(i\sigma_2\Phi(x)^*,\,\Phi(x)\bigr)}
 {\sqrt{\Phi(x)^\dagger\Phi(x)}}=\widehat\Theta(x),
 \qquad \Phi(x)^\dagger\Phi(x)>0.
 \label{eq:ep280-fund-dictionary}
\end{equation}
Thus the scalar supplies a complete frame for the left $SU(2)$
action only away from its zeros.  On a fixed-length lattice,
a simultaneous local gauge transformation $\Omega_x$ and global
right (custodial) transformation $\Gamma$ act as
\begin{align}
 \widehat\Theta_x&\longmapsto
 \Omega_x\widehat\Theta_x\Gamma,\nonumber\\
 \bm n_{\mathscr A,x}&\longmapsto
 \Omega_x\bm n_{\mathscr A,x}\Omega_x^\dagger,
 \quad \Gamma\in SU(2)_{\rm right}\equiv\widetilde{SU(2)}_{\rm global}.
 \label{eq:ep280-double-action}
\end{align}
Accordingly $r_x=\widehat\Theta_x^\dagger
\bm n_{\mathscr A,x}\widehat\Theta_x$ contains the color
axis as \emph{read in the scalar-frame coordinates};
it is
invariant under the local left transformation,
while
under a global custodial right transformation
\begin{equation}
\widehat\Theta_x\mapsto\widehat\Theta_x\Gamma,
 \end{equation}
one has
\begin{equation}
r_x\mapsto\Gamma^\dagger r_x\Gamma .
\end{equation}
The norms $\| R \|$ of the \emph{spatially averaged} $R$, and their susceptibilities, are the reported diagnostics.
The original study explicitly handled a global $\bm n\mapsto-\bm n$ ambiguity by absolute norms and introduced a finite-volume reference subtraction; neither a nonzero local norm $\langle\|r_x\|\rangle$ nor a raw finite-volume $\langle\|R\|\rangle$ would suffice to infer an infinite-volume ordering transition.
Figures 1--15 of Ref.~\cite{Ikeda2024} contain the plaquette and scalar action densities, the corresponding susceptibilities, several norms of $R$, and proposed transition lines, measured in its stated $16^4$ simulation setup.
In particular, the reported additional line is inferred from observables using the CDG reduction rather than from a simple singularity of the original local energy densities.

Using $q_x=\hat\Theta_x^{\dagger}h_{{\mathscr A},x}$ gives the exact relational identity
\begin{equation}
 \boxed{ r_x=q_x\sigma_3q_x^{\dagger} }.
 \label{eq:rxq}
\end{equation}
Thus \textbf{$R$ measures long-distance orientation coherence of a \emph{color-direction} QRF relative to the \emph{fundamental scalar} QRF}, with an induced global right action.
It is not generally the expectation of a strictly local operator of the undeformed original gauge--scalar theory, because \textbf{the optimizing reduction map $U\mapsto n[U]$ is intrinsically nonlocal.}
This provides \textbf{a QRF interpretation of how such a nonlocal reported line can be compatible with the OSFS theorem}.

The QRF interpretation also explains why ``the frame transformation exists'' cannot be the criterion.  For a fixed-length fundamental scalar, $\hat\Theta_x\in SU(2)$ exists on every site, so $q_x$ exists configuration by configuration.  \textbf{The nontrivial information lies in the infrared coherence of $q$ and in the nonlocal way $h_{\mathscr A}[U]$ is determined.}

For $\widehat\Theta_x\in SU(2)$ the scalar frame exists on each site and a relative $q_x$ can always be written once an $h_{\mathscr A}$ prescription is given. Therefore the statement ``$q$ exists in the Higgs region and ceases to exist in the confinement region'' is false for this fixed-length model. The correct possible distinction is a change in relative-frame \emph{coherence, locality, renormalized infrared support, or global order of a specified nonlocal operator}. Even a change in optimal QRF need not coincide with a thermodynamic singularity. The physical mass gap and the endability of fundamental Wilson lines require independent measurement. Consequently the label ``Higgs phase'' used for one side of the nonlocal boundary is operational and need not denote a thermodynamically separate phase.

\section{QRF interpretation of the lattice numerical results}
\label{sec:interpret}

\begin{figure}[h]
\includegraphics[width=45mm]{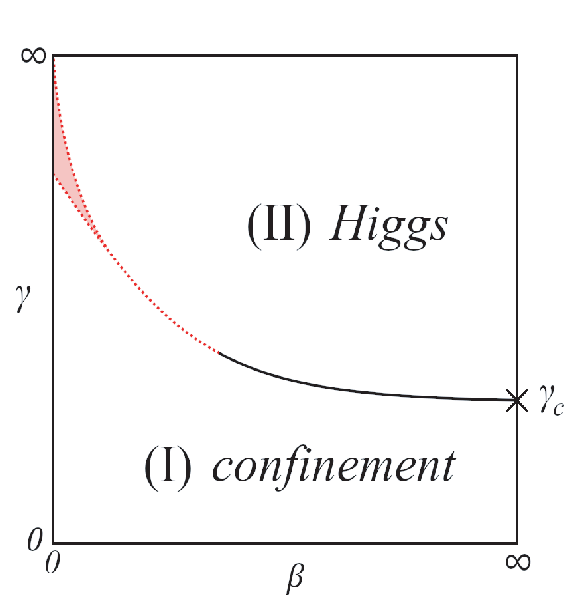}
\caption{
Schematic gauge--fundamental-scalar $SU(2)$ phase diagram adapted/redrawn from Ref.~\cite{Ikeda2024}.
The black continuous
segment depicts the conventional reported weak-coupling boundary;
the red dotted segment depicts the proposed continuation/separation
in the scalar--color diagnostic, and the narrow light-red shaded wedge
illustrates the schematic spread between the dotted curves. No
quantitative confidence interval is assigned to this shading.
The endpoint marked $\gamma_c$ is schematic. Regions (I) and (II)
are labeled ``confinement-like'' and ``Higgs-like'' by the authors'
nonlocal diagnostic; no universal thermodynamic singularity of local
observables follows from this drawing or from finite-volume data.}
\label{fig:phase_diagram2}
\end{figure}

\subsection{Nonlocal transition in the fundamental scalar model: IKKS results}

Ref.~\cite{Ikeda2024} studied the radially fixed $SU(2)$ fundamental gauge--scalar model without gauge fixing.  The standard OSFS result implies that \textbf{confinement-like and Higgs-like regions can belong to one analytically connected thermodynamic phase in the appropriate region of coupling space.}
The simulation nevertheless found \textbf{an additional line detected by the gauge-invariant scalar--color operator $R$, besides the conventional weak-coupling thermodynamic line.}

For each $n=1,2$, with $\|R\|_n$ defined in
Eqs.~\eqref{eq:ep280-Rnorm1} and \eqref{eq:ep280-Rnorm2},
the fixed-length fundamental example of
Ref.~\cite{KondoQRFConfinement2026} uses the schematic
infinite-volume pattern
\begin{align}
 {\rm II}:\quad &
 \langle\|R\|_n\rangle^{\rm sub}>0.
 \label{eq:ep280-fundamental-II}\\
 {\rm I}:\quad &
 \langle\|R\|_n\rangle^{\rm sub}\longrightarrow0,
 \label{eq:ep280-fundamental-I}
\end{align}
The region-II inequality refers to the infinite-volume extrapolation.
The subtraction, susceptibilities and finite-size scaling are
parts of the numerical protocol, rather than consequences of
frame covariance. The exact right/global transformation law
shows that such an alignment measurement can probe
long-distance relational organization and the right symmetry
of the \emph{nonlocally enlarged} description; it is not a
local-gauge-symmetry-breaking expectation value or a new
local order parameter of the original theory.
Fradkin--Shenker analyticity of appropriate local observables
remains consistent with this nonlocal diagnostic.
In addition, dynamical fundamental matter allows fundamental
Wilson lines to end; therefore the $\mathbb Z_2^{(1)}$ electric
center is not an exact symmetry and the above line is not a
center-one-form order parameter.

The reported line is a finite-volume candidate:
\textbf{the analytic OSFS domain concerns suitable \emph{local} gauge-invariant observables and does not exclude either other thermodynamic singularities elsewhere in the coupling plane or nonlocal ordering within an analytically connected region.}
To call this a sharp infinite-volume transition,
one must establish stable volume scaling, a pseudocritical sequence, Binder behavior, reduction-copy stability, and the analyticity or otherwise of local action densities on the same ensembles.
  A nonanalyticity in this sector therefore need not violate OSFS analyticity.  It is best called \textbf{a relative-frame transition} until finite-size scaling establishes whether it is a true nonlocal phase transition or only a sharp crossover.

\textbf{The QRF language also removes a possible misconception: region II is not defined by spontaneous breaking of the local $SU(2)$ redundancy.} \textbf{It is defined by the persistence of a gauge-invariant relation between the gauge-derived and matter-derived frames.}
  The right/global symmetry of the matrix scalar can still be discussed as a genuine global symmetry, but it must not be conflated with the local gauge group.

Finally, the fundamental theory does not possess an exact electric center one-form symmetry in the presence of dynamical fundamental matter.  String breaking and screening are therefore expected.
Consequently, the frame-locking line is not by itself a proof of color confinement in the strong charged-sector sense of Ref.~\cite{KondoQRFConfinement2026}.  One must explicitly include screening and ask whether a nontrivial stable charged class survives the matter channels.

\subsection{IKKS relative-color observables and their QRF meaning}

\label{subsec:QRF-interpretation}

The QRF reading is immediate from Eq.~\eqref{eq:rxq}.  The CDG color-direction field $\bm n[U]$ is an intrinsic gauge-field partial frame, while $\Theta_x$ is a complete matter frame.  The measured quantity $R$ is therefore a volume-averaged relative orientation.  The two regimes may be summarized as
\begin{align}
 {\rm II:}&\quad
 \langle\|R\|\rangle_{\rm sub}>0
 \quad\text{(relative-frame locked)}.
 \label{eq:fundII}\\
 {\rm I:}&\quad
 \langle\|R\|\rangle_{\rm sub}\longrightarrow0
 \quad\text{(relative-frame disordered)},
 \label{eq:fundI}
\end{align}

We propose to examine \textbf{the relative frame correlator} defined by
\begin{align}
 C(x,y) :=\frac12\left\langle \operatorname{tr}
 [r_xr_y] \right\rangle
 = \left\langle \bm r_x \cdot \bm r_y \right\rangle ,
\end{align}
to confirm the QRF interpretation for the IKKS nonlocal transition line.
Our conjecture is that
\begin{align}
 &  \lim_{|x-y|\to \infty}  C(x,y) := C(\infty)
 \nonumber\\
&  \begin{cases}
 \not= 0 & (\text{\textbf{(II) Higgs-like: frame-locking}}) \\
 = 0 & (\text{\textbf{(I) confinement-like: frame disordering}})
 \end{cases} ,
\end{align}
The limiting cases are easily known for any choice of the gauge frame:
\begin{align}
 C(x,y) &=1\not= 0 \quad (\beta=0, \ \gamma=\infty)  ,
\\
 C(x,y) &=0 \quad (\beta=0, \ \gamma=0; \ x\not=y) .
\end{align}
Note that
$C(x,x)=\tfrac12\langle\operatorname{tr}r_x^2\rangle=1$.

For the \emph{specified Wilson-tree} relative-color correlator,
 the exact $\beta=0$ result is
 $C_{n,T}(L_T)=[I_3(\gamma)/I_1(\gamma)]^{L_T}$.
 Thus its long-tree-distance limit vanishes for every finite
 $\gamma\ge0$, as derived in Sec.~\ref{subsec:wilson-vs-cdg}.
For this specified Wilson-tree operator, the exact $\beta=0$
transfer formula gives
for every finite $\gamma$.
\begin{align}
 C_{n,T}(L_T\to\infty)=0 , \quad 0 \le \gamma<\infty.
\end{align}
This operator-specific result is compatible with OSFS
analyticity of suitable local observables, but it is not a consequence
of the OSFS theorem and does not determine the distinct CDG correlator.

Thus \textbf{$R$ measures long-distance orientation coherence of a \emph{color-direction} QRF relative to the \emph{fundamental scalar} QRF}.
This provides \textbf{a QRF interpretation of how such a nonlocal reported line can be compatible with the OSFS theorem},
while leaving open whether the infinite-volume limit has a true singularity, a sharp nonlocal transition, or a crossover, and whether the induced custodial ordering has independent observable spectral consequences.

\textbf{The proposed IKKS/QRF line is therefore interpreted as a
candidate nonlocal relative-frame boundary, potentially distinguishing long-distance CDG--scalar alignment from its absence.}
\textbf{It does not, by itself, separate confinement from the absence of confinement in a charged-sector sense.}
Neither a gauge-invariant vector pole nor an $S_c$-confinement boundary has been determined on the same ensembles. A correspondence between these independent diagnostics remains a testable conjecture.

\begin{widetext}

\begin{center}
\begin{tabular}{@{}lll@{}}
\toprule
 & \parbox[t]{.21\textwidth}{\centering (I) confinement: small-$\gamma$}
 & \parbox[t]{.21\textwidth}{\centering (II) Higgs: large-$\gamma$}\\
\midrule
\parbox[t]{.37\textwidth}{$\bullet$ color neutrality (fundamental matter)}
 & compatible & compatible\\
\parbox[t]{.37\textwidth}{$\bullet$ bare vector stiffness ($\gamma>0$)}
 & nonzero & nonzero\\
\parbox[t]{.37\textwidth}{$\bullet$ gauge-invariant vector pole/mass gap}
 & not determined & not determined\\
\parbox[t]{.37\textwidth}{$\bullet$ fundamental-source screening}
 & dynamical test & dynamical test\\
\midrule
\parbox[t]{.37\textwidth}{$\bullet$ proposed CDG $C_n(\infty)$}
 & zero (candidate) & nonzero (candidate)\\
\parbox[t]{.37\textwidth}{$\bullet$ CDG relative-frame coherence}
 & absent (candidate) & present (candidate)\\
\parbox[t]{.37\textwidth}{$\bullet$ $S_c$-confinement (independent test)}
 & to be measured & to be measured\\
\bottomrule
\end{tabular}
\end{center}

\end{widetext}

\label{qrfA:sec:CDGtest}

\subsection{Fixed-branch CDG expansion at $\beta=0$}

\label{subsec:beta0}

At $\beta=0$, scalar-frame links in the fixed-length ($\rho=v$) fundamental model become independent with one-link weight (see Appendix~\ref{qrfA:app:deriv})
\begin{equation}
 z(\gamma)=\int_{SU(2)}dW\,
 \exp\left[\frac{\gamma}{2}\operatorname{tr}W\right]
 =\frac{2I_1(\gamma)}{\gamma}, \quad \gamma:=\kappa v^2.
 \label{eq:zbeta0}
\end{equation}
The normalized one-link expectation value $t(\gamma)$ is
(see Appendix~\ref{app:latticecalc})
\begin{equation}
 t(\gamma):=
 \left\langle\frac12\operatorname{tr}W\right\rangle
 =\frac{I_2(\gamma)}{I_1(\gamma)}.
 \label{eq:tBessel}
\end{equation}

The CDG reduction functional in the original gauge frame is given by
\begin{align}
\mathcal R[\bm n;U]
&=: \sum_{x,\mu}\left[1-\frac12 {\rm tr}(\bm n_x U_{x,\mu} \bm n_{x+\hat\mu} U_{x,\mu}^\dagger) \right] \nonumber\\
&=\frac14\sum_{x,\mu}\operatorname{tr}\!\left[\left(\bm n_x-U_{x,\mu}\bm n_{x+\hat\mu}U_{x,\mu}^\dagger\right)^2\right].
\label{eq:reduce-original}
\end{align}

For a fixed-length fundamental scalar
$\hat\Theta_x\in SU(2)$, we define the relative frame correlation function $C_n(r)$ for separation $r$ along a lattice axis
\begin{align}
 \bm r_x&=\hat\Theta_x^\dagger n_{x}[U]
       \hat\Theta_x
       =r_x^A\sigma_A,
 \nonumber\\
 C_n(r)&=\left\langle\tfrac12
   \operatorname{tr}(r_xr_{x+r\hat1})\right\rangle
 =\langle\bm r_x\cdot\bm r_{x+r\hat1}\rangle .
 \label{qrfA:eq:Cn}
\end{align}
Here $n_{x}[U]$ is selected by a specified \textbf{CDG reduction algorithm}. Although $r_x$ is assigned to one site, its construction
involves the full gauge configuration $\{ U_{x,\mu} \}$; consequently its long-distance
behavior cannot be inferred from the exact Wilson-tree formula.

In a local branch connected to the uniform solution as $\gamma\to\infty$, we write
$W_{x,\mu}=e^{i\bm a_{x,\mu}\cdot\bm\sigma}$.
$\bm r_x=\bm n_0+\bm\pi_x-\tfrac12|\bm\pi_x|^2\bm n_0+\cdots$
leads to a lattice Poisson equation.
For the four-dimensional Euclidean lattice of size $L$ the Gaussian approximation yields at $\beta=0$
\begin{equation}
 \boxed{C_n(r)_{\beta=0}=1-\frac{8}{\gamma}
 [G_L(0)-G_L(r)]+O(\gamma^{-2}).}
 \label{qrfA:eq:LO}
\end{equation}
This formula is the leading $O(\gamma^{-1})$ \textbf{local-harmonic-branch}
estimate at fixed finite $L$, with a specified zero-mode prescription and
a fixed unit color direction $\bm n_0$. It is not an exact finite-$\gamma$
plateau nor evidence, by itself, for a thermodynamic or nonlocal
transition.
The explicit variation and normalized covariance
calculation are given in Appendix~\ref{qrfA:app:deriv}.
The lattice Green function for separation along a lattice axis is
\begin{align}
 G_L(r)&:=\frac{1}{L^4}\sum_{k\ne0}
 \frac{\cos(k_1r)}{\widehat k^{\,2}},
 \quad \widehat k^{\,2}:=4\sum_{\mu=1}^4\sin^2\!\frac{k_\mu}{2},
 \nonumber\\
 k_\mu&:=2\pi m_\mu/L \qquad (m_\mu=0,\ldots,L-1).
 \label{qrfA:eq:GL}
\end{align}
Its finite-volume expression as a threefold hyperbolic-function
sum and the infinite-volume Bessel integral are derived in Appendix~\ref{qrfA:app:deriv}. In particular,
\begin{align}
 G_\infty(r)&=\int_0^\infty dt\,e^{-8t}
 I_r(2t)I_0(2t)^3,
 \nonumber\\
 G_\infty(0)&=\int_0^\infty dt\,e^{-8t}
 I_0(2t)^4
 =0.154933390231\ldots,\nonumber\\
 G_\infty(r)&\underset{r\to\infty}{\sim}
 \frac{1}{4\pi^2r^2} \to 0.
 \label{qrfA:eq:Ginf}
\end{align}

\begin{figure}[ht]
 \centering

 \includegraphics[width=.98\columnwidth]{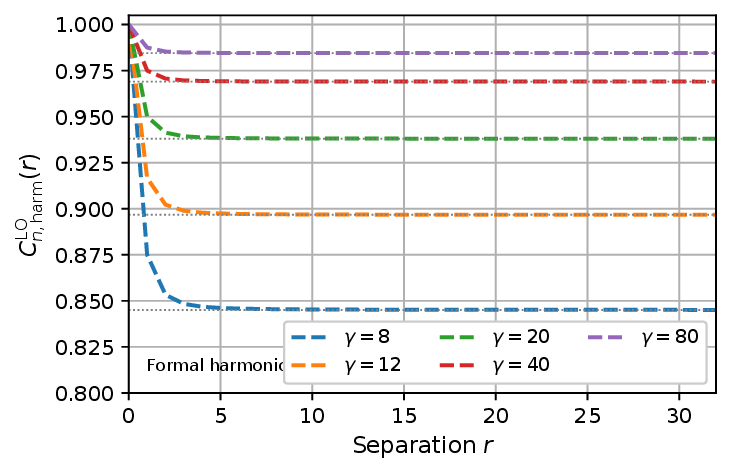}
 \caption{Formal leading harmonic-branch
 estimate only (not Monte Carlo data).
 The curves show $1-8[G_\infty(0)-G_\infty(r)]/\gamma$ for
 the displayed $\gamma$. Dotted horizontal asymptotes are the
 formal leading-order values and do not establish an exact CDG
 plateau or phase transition.}

 \label{qrfA:fig:curves}
\end{figure}

\begin{figure}[ht]
 \centering

 \includegraphics[width=.92\columnwidth]{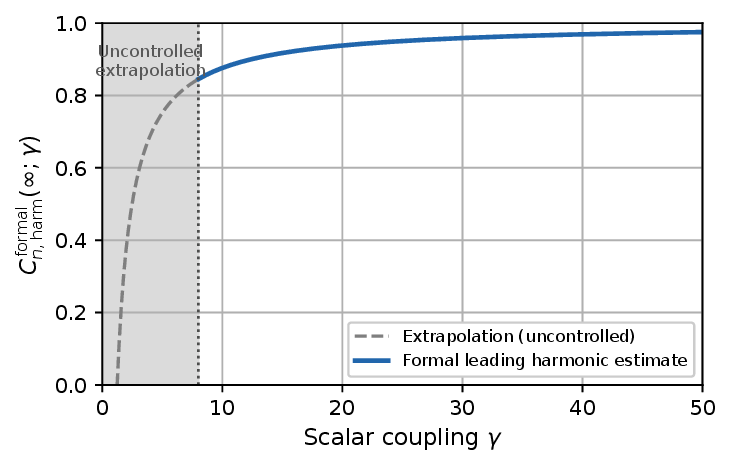}

 \caption{Formal leading harmonic-branch
 extrapolation, $1-1.239467121849/\gamma$, on $0\leq\gamma\leq50$.
 The gray band $\gamma<8$ is outside the illustrative
 large-$\gamma$ display domain; $8$ is \emph{not} a rigorously
 derived threshold. The zero crossing at
 $\gamma=1.239467\ldots$ is \emph{not} a critical point.
 No exact nonperturbative CDG plateau is asserted.}
 \label{qrfA:fig:plateau}
\end{figure}

At fixed finite $L$ and fixed separation $r$, the
local-harmonic-branch expansion gives Eq.~\eqref{qrfA:eq:LO}.
A \emph{formal} infinite-volume leading coefficient is obtained
by taking the corresponding infinite-volume Green-function limit:
\begin{equation}
 C_{n,\mathrm{harm}}^{\mathrm{formal}}(\infty;\gamma)
 :=1-\frac{8G_\infty(0)}{\gamma}
 =1-\frac{1.239467121849\ldots}{\gamma}.
 \label{qrfA:eq:plateau}
\end{equation}
This notation denotes an extrapolation of the leading harmonic
coefficient.
However, it {does not establish} an exact
$\lim_{r\to\infty}\lim_{L\to\infty}C_n(r)$ or an $O(\gamma^{-2})$ error bound uniform in $L,r$.
An interchange of the large-$\gamma\to\infty$,
thermodynamic $L\to\infty$, and separation limits $r\to\infty$ requires additional estimates.
Figures
\ref{qrfA:fig:curves} and \ref{qrfA:fig:plateau} display only the
formal leading harmonic-branch estimate.

At small hopping $\gamma \ll 1$ the normalized \textbf{one-link adjoint transfer coefficient} is exactly
\begin{equation}
 \lambda_1(\gamma)=\frac{I_3(\gamma)}{I_1(\gamma)}
 =\frac{\gamma^2}{24}-\frac{\gamma^4}{384}
 +O(\gamma^6).
 \label{qrfA:eq:lambda}
\end{equation}
A tentative \textbf{shortest-path hopping estimate} for the \emph{CDG} observable has the proposed leading form
\begin{equation}
 C_n(r)\stackrel{\rm ansatz}{\approx}K_r^{(0)}
 \lambda_1(\gamma)^r
 =K_r^{(0)}\left(\frac{\gamma^2}{24}\right)^r
 +O(\gamma^{2r+2}),
 \label{qrfA:eq:small}
\end{equation}
where $K_r^{(0)}$ is a transported color-direction correlation
computed with $\gamma=0$ Haar links and the same reduction
prescription in Appendix~\ref{qrfA:app:deriv}.
The coefficient $K_r^{(0)}$ depends on the specified reduction protocol.
Unlike the Wilson-tree transfer identity, this CDG expansion
requires control of the nonlocal dependence of the minimizing
frame on the links; Eq.~\eqref{qrfA:eq:small} should therefore
be used as a short-distance hopping estimate, not as a general
all-distance identity.

The two asymptotic calculations are consistent with strong
alignment at sufficiently large $\gamma$ and short-range
correlations at small $\gamma$. They do \emph{not} establish a
nonzero exact $C_n(\infty)$ for finite $\gamma$, a critical
coupling, or a thermodynamic singularity. In particular, a
term of order $\gamma^{-2}$ involves nonlinear reduction-field
responses and may fail to be uniform at long distances in four
dimensions. A copy-controlled, volume-scaled comparison with
local physical observables remains necessary.

\subsection{Wilson-tree control: exact $\beta=0$ and Gaussian benchmarks}

\label{subsec:wilson-vs-cdg}

We now consider
the \emph{Wilson-rooted} frame instead of the reduction-selected CDG direction.
The Wilson-rooted construction and its
root-fiber transformation law have already been given in
Sec.~\ref{subsec:wilsontreecontrol} and Appendix~\ref{app:wilsontree};
we therefore use them without repeating their general derivation.

For the fixed-length scalar let
$\widetilde W_{x,\mu}=\widehat\Theta_x^\dagger
U_{x,\mu}\widehat\Theta_{x+\hat\mu}$ be the scalar-frame link,
and let $P_x$ be the unique rooted-tree path. With the link products
oriented consistently with Eq.~\eqref{eq:propositiontree},
\begin{align}
q_{T,x} &=\widetilde W(P_x)^\dagger q_{T,x_0},\quad
 \widetilde W(P_x)^\dagger
 =\widetilde W_{x_{m-1},x_m}^\dagger
 \cdots\widetilde W_{x_0,x_1}^\dagger,\nonumber\\
 r_{T,x}&=q_{T,x} \bm n_0q_{T,x}^\dagger
 =\widetilde W(P_x)^\dagger r_{T,x_0}\widetilde W(P_x),
 \label{eq:WilsonAdjointPath}
\end{align}
where $r_{T,x_0}=q_{T,x_0} \bm n_0q_{T,x_0}^\dagger$,
$\bm n_0^2=\bm1$, and the root axis $\bm n_0$ obeys the
covariance prescription in Eq.~\eqref{eq:rootcovaxis}.
Here $\bm n_0=\sigma_3$ is permitted only in a \emph{fixed root
trivialization}; a numerical axis held fixed under unrestricted
root gauge transformations would not give a gauge-invariant
observable.
On each forward tree edge $x\to x+\hat\mu$ the unambiguous
one-step relation is
\begin{equation}\label{eq:tree-step-A4}
 q_{T,x+\hat\mu}=\widetilde W_{x,\mu}^\dagger q_{T,x},
 \quad \widetilde W_{x,\mu}=\widehat\Theta_x^\dagger
 U_{x,\mu}\widehat\Theta_{x+\hat\mu}.
\end{equation}
This fixes the ordering in the tree identities in
Eqs.~\eqref{eq:propositiontree} and \eqref{eq:treeid}.

Thus the relative-frame correlator on the tree
\begin{equation}
 C_{n,T}(x,y):=\frac12\left\langle
 \operatorname{tr}\bigl(r_{T,x}r_{T,y}\bigr)
 \right\rangle
 \label{eq:WilsonAdjointC}
\end{equation}
is well defined for that root/boundary protocol. Its adjoint
color-axis insertion distinguishes it from the fundamental
scalar--string correlator $C_q^{(T)}$ in
Eq.~\eqref{eq:propositiontree}; neither is the CDG $C_n$.

Define the one-link transfer map acting on a traceless Hermitian matrix $X=X^A\sigma_A$ by
\begin{equation}
 {\cal T}_\gamma[X]
 :={1\over z(\gamma)}
 \int d\mu(W)\,e^{\frac\gamma2\mathrm{tr}W}\,
 W^\dagger X W
 \label{eq:T}
\end{equation}
Since the weight is a class function, Schur's lemma implies that, in the adjoint sector,
\begin{equation}
 {\cal T}_\gamma[X]=\lambda_1(\gamma)X.
 \label{eq:eigen}
\end{equation}
The character of the spin-one representation of $SU(2)$ is
\begin{equation}
 \chi_1(\theta)=1+2\cos2\theta
 =4\cos^2\theta-1,
 \label{eq:chi1}
\end{equation}
and, since the dimension of the adjoint representation is three,
\begin{equation}
 3\lambda_1(\gamma)
 ={1\over z(\gamma)}
 \int d\mu(W)\,e^{\frac\gamma2\mathrm{tr}W}\chi_1(\theta).
 \label{eq:lambda-char}
\end{equation}
Using the integral representation of Bessel functions gives
\begin{equation}
 \int d\mu(W)\,e^{\gamma\cos\theta}\chi_1(\theta)
 =I_2(\gamma)-I_4(\gamma)
 ={6I_3(\gamma)\over\gamma} .
 \label{eq:character-integral}
\end{equation}
Substituting Eq.~(\ref{eq:z}), we obtain the eigenvalue
\begin{equation}
 \lambda_1(\gamma)={I_3(\gamma)\over I_1(\gamma)}>0.
 \label{eq:lambda}
\end{equation}
The coefficient is the same one-link adjoint ratio appearing in
Eq.~\eqref{qrfA:eq:lambda}, but its \emph{application to the full
two-point operator} is exact only for the specified tree frame.

Let $L_T(x,y)=|T_{xy}|$ be the \emph{tree distance}, i.e., the number of links on the spanning tree connecting $x$ and $y$, which is not necessarily equal to the Euclidean distance.
Since Eq.~(\ref{eq:eigen}) can be applied to each independent link,
repeated conditional integration over the independent links of the unique tree path $T_{xy}$ gives the exact expression independent of the spacetime dimension
\begin{align}
 &
 C_{n,T}(x,y)\big|_{\beta=0}
 =\lambda_1(\gamma)^{L_T(x,y)}
 =\left[\frac{I_3(\gamma)}{I_1(\gamma)}\right]^{L_T(x,y)},
 \label{eq:WilsonAdjointExact}\\
& C_q^{(T)}(x,y)\big|_{\beta=0}
 =\left[\frac{I_2(\gamma)}{I_1(\gamma)}\right]^{L_T(x,y)},
 \label{eq:WilsonFundExact}
\end{align}
The second expression is the
fundamental representation counterpart of the first and agrees with
the one-link ratio $t(\gamma)$ in Eq.~\eqref{eq:tBessel}.
The distinct Bessel ratios reflect two distinct representation
channels of the \emph{same} prescribed Wilson tree.
For finite $\gamma$,
\begin{equation}
 0\le \lambda_1(\gamma)={I_3(\gamma)\over I_1(\gamma)}<1,
\label{eq:bound}
\end{equation}
and hence
\begin{equation}
 \boxed{
 \lim_{L_T\to\infty}C_{n,T}(x,y)\big|_{\beta=0}=0
 \quad(\beta=0,\;\gamma<\infty; \ x\not=y)
 }.
 \label{eq:beta0zero}
\end{equation}
Defining \textbf{the relative-frame coherence length } $\xi_{\rm rel}$ in terms of the distance along the spanning tree by
\begin{equation}
 C_{n,T}(L_T)=e^{-L_T/\xi_{\rm rel}}
 \label{eq:xi-def}
\end{equation}
we have
\begin{equation}
 \boxed{
 \xi_{\rm rel}(\gamma)
 ={1\over-\log[\lambda_1(\gamma)]}
 }.
 \label{eq:xi-exact}
\end{equation}
At large hopping $\gamma$, $\xi_{\rm rel}$ grows but remains finite at every finite $\gamma$ along $\beta=0$.
\begin{align}
 \lambda_1(\gamma)
 &=1-\frac{4}{\gamma}+\frac{6}{\gamma^2}
   -\frac{3}{2\gamma^3}+O(\gamma^{-4}),
 \nonumber\\
 \xi_{\rm rel}(\gamma)
 &=\frac{\gamma}{4}-\frac18
   +\frac{13}{96\gamma}+O(\gamma^{-2}).
 \label{eq:WilsonAdjointLarge}
\end{align}

At $\gamma=0$ the correlation length vanishes $\xi_{\rm rel}=0$. More explicitly,
\begin{align}
 \lambda_1(\gamma)
 &=\frac{\gamma^2}{24}-\frac{\gamma^4}{384}
   +\frac{\gamma^6}{5760}+O(\gamma^8).
 \label{eq:WilsonAdjointSmall}
\end{align}
Therefore, we have
\begin{equation}
 C_{n,T}(L_T)
 =\left[{\gamma^2\over24}
 +O(\gamma^4)\right]^{L_T},
 \label{eq:smallC}
\end{equation}
and \textbf{relative-frame decoherence is extremely strong in the small-$\gamma$ confinement-like region}.
In particular, at leading order, extending the path by one link contributes an additional factor of $\gamma^2/24$.

In contrast, the strict fixed-path $\gamma=\infty$ endpoint has
$\widetilde W_\ell=\bm1$ and $C_{n,T}=1$. Consequently, \textbf{the thermodynamic/long-distance limit and the $\gamma \to \infty$ limit do not commute}:
\begin{equation}
 \lim_{\gamma\to\infty}\lim_{L_T\to\infty}C_{n,T}=0,
 \qquad
 \lim_{L_T\to\infty}\lim_{\gamma\to\infty}C_{n,T}=1
 \quad(\beta=0).
 \label{eq:WilsonAdjointNoncommute}
\end{equation}
The same scalar-frame Haar transformation acts on \emph{all}
positively oriented lattice links, not only tree edges. Hence, with
normalized Haar measure,
\begin{align}\label{eq:free-energy-B2}
 \frac{Z(0,\gamma)}{Z(0,0)}
 &=\left[\frac{2I_1(\gamma)}{\gamma}\right]^{N_\ell},
 \qquad N_\ell=D V,
\end{align}
which implies the free-energy difference
\begin{align}
 f(0,\gamma)-f(0,0)
 &=-D\log\!\left[\frac{2I_1(\gamma)}{\gamma}\right],
\end{align}
where
\begin{align}
 f(\beta,\gamma):=-\frac{\log Z(\beta,\gamma)}{V}.
\end{align}
Here $V=N_{\rm site}$ and a periodic four-dimensional lattice has
$N_\ell=4V$, so $f(0,\gamma)-f(0,0)=-4\log[2I_1(\gamma)/\gamma]$.
The value at $\gamma=0$ is defined by continuity, $z(0)=1$;
for any finite real $\gamma$, $z(\gamma)>0$ and this free-energy difference is real analytic. It thus coexists with the exact
$C_{n,T}\sim[I_3(\gamma)/I_1(\gamma)]^{L_T}$ path decay.
This is a theorem about the fixed-length \emph{bulk} partition function and the separately defined Wilson-tree observable, and places no analyticity constraint on the reduction-selected CDG
$C_n$ in its nonlocal thermodynamic limit.

This exact noncommutativity is a property of the chosen Wilson-tree operator. In particular, at $\beta=0$ it has
\emph{no finite-$\gamma$ nonlocal ordering point}, and the analyticity
of the local partition function is not a proof of an analogous
statement for a reduction-selected CDG observable.

As $\gamma\to\infty$, the link variables expressed in the scalar frame are fixed to
$
 W_{x,\mu}\longrightarrow {\bf1}
$
Therefore, for any path,
$
 W(P_x)={\bf1}
$, $
 q_x=q_\star
$, $
 r_x=r_\star
$
and
\begin{equation}
 C_n(x,y)=1
 \qquad(\gamma=\infty).
 \label{eq:Cinfty}
\end{equation}
Thus, complete relative-frame locking holds exactly at the Higgs endpoint.

Next consider four dimensions, with $\gamma$ sufficiently large but finite and $\beta$ also finite.
For fixed finite $\beta$ and large $\gamma$,
write the scalar-frame links as
$
 W_{x,\mu}=\exp\{i a_{x,\mu}^A\sigma_A\}
$.
The quadratic approximation to the four-dimensional Euclidean action then reads
\begin{equation}
 S_{\rm G}
 ={1\over2}\sum_{x,A}
 \left[
 \gamma\sum_\mu(a_{x,\mu}^A)^2
 +\beta\sum_{\mu<\nu}
 (\Delta_\mu a_{x,\nu}^A-\Delta_\nu a_{x,\mu}^A)^2
 \right].
 \label{eq:gaussian-action}
\end{equation}
The parameter $\gamma$ provides a mass term for the gauge-invariant vector mode.

Consider a straight Wilson path of length $L$ in the $\mu=4$ direction.
The leading variance, per color component, of the integrated link rotation
$
\Omega^A_L=\sum_{s=0}^{L-1}a_{x+s\hat4,4}^A
$
is, for large $L$,
\begin{align}
\langle(\Omega_L^A)^2\rangle=&LJ_3(\beta,\gamma)+O(1), \\
 J_3(\beta,\gamma)
 =:&\int_{[-\pi,\pi]^3}\frac{d^3p}{(2\pi)^3}\,
 \frac{1}{\gamma+4\beta\displaystyle
 \sum_{j=1}^3\sin^2(p_j/2)}.
 \label{eq:WilsonAdjointJ3}
\end{align}
A rotation $\exp(i\Omega^A\sigma_A)$ in the fundamental representation of $SU(2)$ rotates an adjoint vector through an angle $2|\bm\Omega|$.

The leading Gaussian cumulant of the two transverse adjoint
rotations of the \emph{fixed straight Wilson-tree axis} yields
\begin{equation}\label{eq:kappa}
 \log C_{n,T}^{\mathrm{Gauss}}(L)
 =-4LJ_3(\beta,\gamma)+O(L\gamma^{-2})+O(1),
\end{equation}
where the $O(L\gamma^{-2})$ term denotes the expected order
\emph{within the large-$\gamma$ expansion}, not a proven
remainder estimate uniform in arbitrary $L$. For finite
$\beta\geq0$ and $\gamma>0$ the Gaussian coefficient obeys
\begin{equation}\label{eq:Jpositive}
 J_3(\beta,\gamma)>0.
\end{equation}
At $\beta=0$, $4J_3=4/\gamma$ agrees with the
large-$\gamma$ expansion of the \emph{exact tree} Bessel ratio.
The decay prediction concerns $C_{n,T}$, not the
reduction-selected $C_n(r)$; positivity of the leading exponent
alone does not prove its exact infinite-length limit for all
finite couplings. In particular no unconditional boxed
all-$\beta,\gamma$ limit is inferred.

This does not mean that Higgs-like physics disappears.
Although each link satisfies $W_{x,\mu}\simeq{\bf1}$, a Wilson line is a product of many ``nearly identity matrices.''
Infinitesimal adjoint rotations diffuse in proportion to the path length, producing perimeter-type self-decoherence in the raw Wilson-line correlator itself.

The $4J_3$ term reduces to $4/\gamma$ at $\beta=0$,
in agreement with the leading exponent of the exact
Bessel result in Eq.~\eqref{eq:WilsonAdjointLarge}.
Thus even when individual scalar-frame links lie close to the
identity, their small rotations accumulate along a long
Wilson line and can produce a perimeter-type decay in a
Higgs-like regime. Equation~\eqref{eq:kappa}
is a controlled \emph{leading large-$\gamma$ coefficient for the
specified straight path}; it is \emph{not} an all-coupling proof
of $\lim_{L\to\infty}C_{n,T}(L)=0$ at every finite
$\beta,\gamma$. Uniform control of the neglected terms in $L$
and a prescribed line-renormalization procedure would be needed
for such a conclusion.

The CDG local-branch large-$\gamma$ approximation
in Eq.~\eqref{qrfA:eq:LO} approaches the \emph{perturbative}
plateau in Eq.~\eqref{qrfA:eq:plateau}, while the Wilson-tree
observable has the exact $\beta=0$ path-length decay in
Eq.~\eqref{eq:WilsonAdjointExact}. These results are compatible:
the CDG direction is selected by a nonlocal reduction of the full link configuration, whereas the Wilson prescription transports a reference axis along a fixed tree. Neither operator identity nor equality of infrared limits follows between them.

On the $\beta=\infty$ boundary, flat links on a simply connected
region can be gauged to unity, and the fixed-length theory reduces
to an $O(4)$ spin model. An ordered regime with nonzero long-range
orientation is possible there at sufficiently large hopping.
Nontrivial global holonomies on a torus and the change of limits
must be treated separately; this boundary result cannot simply
be continued to finite $\beta$.

The $\beta=\infty$ spin-model endpoint therefore cannot establish the behavior of the separately defined CDG and Wilson-tree correlators at any finite $\beta$. In particular, the Gaussian line coefficient of Eq.~\eqref{eq:kappa} is restricted to the Wilson-tree diagnostic.

\subsection{Complementarity reinterpreted}

  In a fundamental theory, Higgs-like and confinement-like regions may remain in one thermodynamic phase while the \emph{optimal relational description} changes.  The matter frame can become a highly coherent local reference and the massive relational link becomes the economical variable.  On the confinement-like side, a Wilson-line or gauge-field-derived frame can be more natural, and long-distance orientation may be disordered or screen through extended gauge structure.  The crossover or nonlocal transition is then a change in relational organization, not necessarily a change of local thermodynamic phase.

The alignment diagnostics resolve \emph{which} independently
constructed color frames exhibit long-distance relative order,
and can be combined with measurements of scalar defects and
nontrivial Wilson/center sectors. They do not by themselves
decide the energetics of an asymptotically isolated charged
sector. In the fundamental model, the available screening
channels must first be accounted for when defining stable
charged classes; any remaining finite-energy, physically
accessible nontrivial class requires an independent
Gauss-law/boundary and isolation-energy analysis as in
Ref.~\cite{KondoQRFConfinement2026}.

The implication for \emph{pure} four-dimensional Yang--Mills
theory is not that every intrinsic gauge-derived color frame
must decohere for confinement to hold. Instead one may choose
an admissible partial direction
$\bm n_{\mathscr A}\in\mathcal N[\mathscr A]$ (where
$\mathcal N[\mathscr A]$ denotes a specified set of
reduction-selected candidate directions), or an admissible
complete dressing $h[\mathscr A]$, and test its infrared
correlation and defect statistics under an explicit
copy/path/boundary protocol. Disorder of a particular selected
frame is neither a necessary nor a sufficient condition for
charged-sector exclusion. The independent finite-energy,
long-distance charged-sector test is the actual confinement
criterion; frame ordering remains a potentially informative
observable or ingredient of a dynamical mechanism.

\section{Predictions with an unfixed radial mode}
\label{sec:radialpred}

The published fixed-length measurements ~\cite{Ikeda2024,ShibataKondo2024} can be extended
to a fluctuating radius by comparing the threshold distribution
$P_\epsilon=\Pr(\rho_x<\epsilon)$ at several nonzero
$\epsilon$, the regular radius-weighted correlator $C_\Psi$,
and a conditional normalized-frame correlator at the same
$(\beta,\kappa,\lambda,V)$. Exact zeros have probability zero
at any given finite-lattice site for a continuous radial density,
but a threshold event does not have probability zero.

For a fundamental scalar, the normalized matter frame exists only where $\rho_x\neq0$.  The relative-frame observable should therefore be accompanied by the scalar-zero probability $P_\epsilon$, the radial susceptibility, the regular correlator $C_{\Psi}$, and a conditional orientation correlator evaluated only on sites/paths with $\rho$ above a fixed threshold.  If the fixed-length transition is genuinely a frame-locking phenomenon, the long-distance locking line should persist for sufficiently stiff radial potentials but can broaden, move, terminate, or convert into a crossover when scalar zeros proliferate.  A distinct radial critical line can also appear if the scalar potential itself undergoes a thermodynamic transition.

The exact $\beta=0$ formulas of Sec.~\ref{subsec:beta0} provide benchmarks.  With a sharply peaked radial distribution, Eq.~\eqref{eq:Cqradexact} reduces to the fixed-length Bessel ratio.  With broad radial fluctuations, rare small-$\rho$ links strongly reduce the product and can shorten the normalized frame coherence even when the mean radius remains appreciable.  The regular $C_\Psi$ is less singular because it weights such configurations by endpoint radii.  Comparing these two observables separates failure of the normalized frame from ordinary loss of scalar propagation.

With $\rho_x$ dynamical, the two kinds of QRF correlator in Eq.~\eqref{eq:CqCPsi} separate orientation coherence from scalar-amplitude physics. We propose to measure
\begin{align}\label{eq:smallrho}
 P_\epsilon&=\left\langle\frac1{N_s}\sum_x\mathbf1_{\rho_x<\epsilon}\right\rangle,\\
 & C_q^T,\quad C_\Psi^T,\quad\langle\rho_x\rho_y\rangle_{\rm conn}.
\end{align}
Report $P_\epsilon$ as a function of the \emph{dimensionless,
renormalized} threshold, the lattice spacing, and any smoothing scale.
For a continuous fixed-site radial probability, exact zeros have
zero probability. An observed threshold-crossing cluster cannot
be assigned a topological winding charge without an independent
oriented-sphere measurement. Both definitions must be retained
before interpreting ``defect density'' in a continuum limit.

A small-$\rho$ density and its clustering could lead to practical frame failures without turning a pointwise measure-zero zero into a theorem about phase boundaries. The lattice radial kinetic interactions also shift the effective coupling and may change or remove the observed nonlocal crossover/critical line. Which possibility occurs is a numerical prediction to test, not an automatic consequence of Fradkin--Shenker continuity. Finite-volume and order-of-limits care is essential, and the relation to the FMS scalar and vector spectra should be checked through invariant interpolating operators.

\section{Other approaches to defect-supported and nonlocal distinctions inside confinement--Higgs continuity}
\label{sec:defects}

The confinement--Higgs correspondence can hold for local bulk observables while a nonlocal object or a defect world volume supports additional sharp structure.

Hayata, Hidaka, and D.~Kondo \cite{HayataHidakaKondo2025} studied a $(3+1)$-dimensional $U(1)_{\rm gauge}\times U(1)_{\rm global}$ model with superfluid vortices. Their simulations find $\mathbb Z_2$ flavor order \emph{on a vortex} in the weak-coupling Higgs regime and its absence in the strong-coupling confinement regime, with a second-order transition of two-dimensional Ising type.
Bulk analyticity or crossover behavior of local observables does not by itself preclude a \emph{defect-localized} transition.

A QRF interpretation is given
as a proposal. Around a vortex, gauge transport $h_{\mathscr A}$ and the matter/superfluid orientation define relative holonomy data. Restricting that data to the vortex world sheet, one can define a framed defect observable $\mathcal O_{\mathcal V}$ carrying a genuine $\mathbb Z_2$ flavor quantum number. A candidate correlator is
\begin{align}\label{eq:vortex}
 C_{\mathcal V}(u,v) =\left\langle\mathcal O_{\mathcal V}(u)
 \mathcal O_{\mathcal V}(v)\right\rangle_{\mathcal V}, \quad
  u,v\in\mathcal V.
\end{align}
To reproduce Ref.~\cite{HayataHidakaKondo2025} rather than merely rename its order parameter, one must construct $\mathcal O_{\mathcal V}$ explicitly from the microscopic fields and the known $\mathbb Z_2$ transformation, prove its frame covariance and residual gauge invariance, and match its scaling dimension and Ising exponents. The general QRF geometry motivates these steps but does not establish them automatically.

The defect operator $\mathcal O_{\mathcal V}$ is therefore a
candidate, not a microscopically constructed quantity in this paper;
its $\mathbb Z_2$ charge and Ising exponent must be demonstrated in
the underlying model.

Several related works fit naturally into the same organization.

Hayashi \cite{Hayashi2024} showed that fractional Aharonov--Bohm phases around vortices can match smoothly between confining and Higgs regimes in theories with fundamental matter, supporting continuity rather than a mandatory bulk transition.  In QRF language the Aharonov--Bohm phase is a relational holonomy around the defect; its smooth matching says that the same topological relational datum can be transported continuously between the two bulk descriptions.

The vortex-holonomy studies of Cherman \emph{et al.} \cite{Cherman2020} and Hayashi \cite{Hayashi2024} also make clear that claims of a necessary nonanalyticity of a vortex Aharonov--Bohm observable are \emph{model-dependent}. In relevant lattice systems matching Aharonov--Bohm phases are smooth across the confinement--Higgs crossover, whereas protected flavor order on a vortex can still distinguish defect sectors. The two findings concern different observables and should not be conflated. A useful common test is whether a relative-frame holonomy, with the correct boundary and counterterm prescription, changes nonanalytically in the same defect sector as the measured flavor order.

Wenzel \emph{et al.} found in the three-dimensional compact $U(1)$ lattice Higgs model with a fluctuating Higgs amplitude a Kertesz line across which vortex proliferation changes while thermodynamic quantities remain nonsingular beyond the endpoint of a first-order line \cite{Wenzel2005}.  This is a close precedent for separating defect/percolation structure from a local thermodynamic singularity.  The relative-frame line of a fundamental $SU(2)$ model may or may not be a Kertesz-type line, but this is a concrete hypothesis that can be tested numerically.

Greensite and Matsuyama's nonlocal custodial diagnostic provides another precedent: a global-symmetry-related nonlocal transition can occur without contradicting the OS theorem \cite{GreensiteMatsuyama2018}.

Shimada and Yamamoto compared Polyakov, $\,'$t Hooft,
and Aharonov--Bohm nonlocal observables in a \emph{charge-two
Abelian Higgs lattice model} \cite{ShimadaYamamoto2025}.
Their results demonstrate a useful contrasting nonlocal operator
protocol, but are \emph{not} a simulation or direct numerical
verification of the $SU(2)$ fundamental CDG color direction.
The operator algebras and the representation of the dynamical
matter must therefore be kept distinct.

\section{Conclusion and discussion}
\label{sec:conclusion}

We have developed \textbf{a single QRF framework for the confinement--Higgs relation whose central dynamical object is the relative orientation between a confinement-oriented gauge-field frame and a Higgs-oriented matter frame.}
The exact identities show how the two descriptions are related; the long-distance correlators and charged-sector tests determine what the relation means dynamically.
Only the separately defined frame-sensitive observables can be used to address alignment of two independently constructed frames.
The matching of such alignment to confinement physics remains a falsifiable physical proposal.

The gauge-invariant dressing identities used in this construction are not claimed as new by themselves.
\textbf{The distinct contribution is the simultaneous comparison of independently constructed gauge-derived and matter-derived frames, together with a same-ensemble program that tests their relative orientation against charged-sector energetics, center sectors, defects, and gauge-invariant spectral observables.}

$\bullet$ For a fundamental $SU(2)$ scalar, a nonzero field gives a complete local \emph{matter} frame.
The gauge-invariant link of the BEH construction is exactly the gauge link in this scalar QRF, and the Ikeda--Kato--Kondo--Shibata \cite{Ikeda2024} scalar--color variable $R$ is exactly a gauge-derived color direction read in the scalar frame.
Their new line is therefore naturally interpreted as \textbf{a relative-frame locking line}.
Because the diagnostic contains a nonlocally determined gauge-field frame and because dynamical fundamental matter removes exact center one-form symmetry, this interpretation is compatible with Fradkin--Shenker continuity.
The line need not be a local thermodynamic phase boundary.

$\bullet$ Restoring the radial scalar adds another layer.  The scalar amplitude weights the quality of the Higgs QRF, scalar zeros are local failures of the normalized frame, and the unnormalized relational scalar provides a regular observable.  This leads to a concrete program for extending the fixed-length simulations.  The same conceptual structure also explains how a defect-supported transition, such as the vortex-worldsheet Ising transition of Hayata--Hidaka--Kondo, can coexist with a smooth bulk confinement--Higgs crossover.

$\bullet$ To establish the physical claims suggested by the two lattice studies \cite{Ikeda2024,ShibataKondo2024}, several tests are still required.  Most importantly, the new boundary must be characterized by finite-size scaling and by an observable set larger than the original scalar--color magnitude.  The frame prescription should be varied within a controlled admissible class (for example CDG reduction, Laplacian-type frames, and specified Wilson/tree frames) to determine which infrared features are universal.  Reduction-condition copy dependence should be studied by best-of-$N$ or weighted-copy extrapolations rather than by selecting an uncontrolled local minimum.

In the fundamental theory, the relative-frame coherence length should be compared directly with screening/string-breaking scales and gauge-invariant vector/scalar masses; this will decide whether the new line is a genuine nonlocal transition, a Kertesz-type line, or a crossover.

Finally, the same measurements should be repeated with a fluctuating radial mode, including scalar-zero statistics and the regular unnormalized relational correlator.

$\bullet$ These tests give clear falsification criteria.  If the location or scaling of the proposed relative-frame boundary changes arbitrarily under admissible frame prescriptions, the boundary is not a physical QRF feature.  If frame locking or disorder has no reproducible relation to charged-sector energetics, masses, defects, or residual-sector confinement, it is only a descriptive correlation.  Conversely, correlated continuum scaling of the relational coherence length, charged-sector energy, mass spectrum, and defect observables would establish that the lattice lines are manifestations of a robust gauge-invariant relational structure.  The resulting picture would not replace the OSFS theorem or 't Hooft complementarity; it would sharpen them by identifying \emph{what can change physically and relationally even when local thermodynamic analyticity persists}.

\section*{Acknowledgments}
This work was supported by Grant-in-Aid for Scientific Research, JSPS KAKENHI Grant
Number (C) No.23K03406.
During preparation of this manuscript, the author used ChatGPT (OpenAI; GPT-6, GPT-5.6 Sol and GPT-5.5) for language editing, restructuring selected passages, and preliminary consistency checks of derivations and bibliography.
The author independently rederived or verified all scientific suggestions against the relevant calculations and cited sources, reviewed all AI-assisted text, and takes full responsibility for the accuracy, citations, scientific claims, and conclusions of the manuscript.

\section*{Data Availability}
No new numerical data were generated in this theoretical study. Numerical results discussed in the text are taken from the cited literature.

\appendix

\section{Wilson-rooted frames as a control prescription}
\label{app:wilsontree}

This appendix records the Wilson-rooted construction that is deliberately compressed in the main text. It is included as a control case for path dependence and root covariance, not as a substitute for the CDG reduction observable.

\textbf{A rooted spanning tree} $T$ is a connected, cycle-free subgraph containing all lattice sites and determines a unique path $P_x:x_0\to x$. If $P_x$ and $P_y$ share a root segment, that segment cancels in $U(P_x)U(P_y)^\dagger$, leaving the unique tree path $T_{xy}$. Off-tree links remain in the gauge dynamics and closed non-tree cycles still detect curvature.

\begin{figure}[h]
 \centering
 \includegraphics[width=.96\columnwidth]{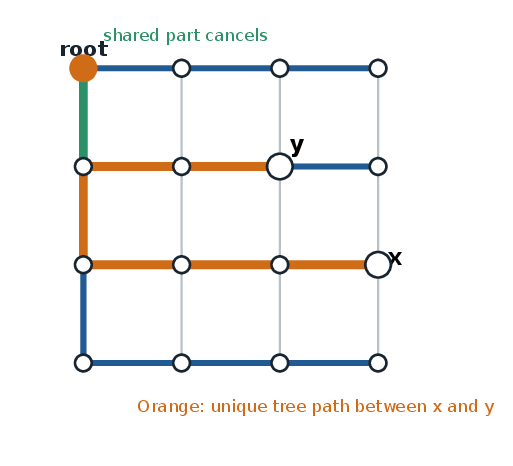}
 \caption{Rooted spanning tree used only as a control prescription. The common root segment cancels between the two root transports, while the remaining segments form the unique tree path $T_{xy}$. Off-tree links are not removed from the gauge dynamics.}
 \label{fig:ep284_wilsontree_appendix}
\end{figure}

Because $h_{\mathscr A,T}(x)=U(P_x)$ transforms at both endpoints, the common root fiber is part of the operator prescription. A fixed numerical color axis at the root is not covariant under unrestricted root transformations; either the root axis must transform as in Eq.~\eqref{eq:rootcovaxis} or a physical root/boundary convention must be specified. Changing the tree changes the nonlocal operator, not merely its gauge representation.

For a straight radial Wilson prescription in a smooth star-shaped patch one obtains the Fock--Schwinger-type condition
\begin{equation}
 (x-x_0)^\mu\mathscr A^{h_{\mathscr A}}_\mu(x)=0,
\end{equation}
and the corresponding connection can be represented in terms of curvature parallel transported to the common root fiber. This simplification never implies $\mathscr F=0$: different paths are related by Wilson holonomy, and closed loops retain the curvature information. These facts are precisely why the Wilson-rooted correlator in Eq.~\eqref{eq:propositiontree} is a useful control observable but cannot be identified with a reduction-derived CDG correlator.

\section{Lattice hopping and one-link calculations}
\label{app:latticecalc}

\subsection{Character expansion at fixed length}

At $\beta=0$ the one-link Boltzmann factor admits an $SU(2)$ character expansion
\begin{equation}
 e^{\frac{\gamma}{2}\operatorname{tr}W}
 =\sum_{j=0,\frac12,1,\ldots}a_j(\gamma)\chi_j(W),
 \label{eq:character}
\end{equation}
with coefficients
\begin{equation}
 a_j(\gamma)=\frac{2(2j+1)}{\gamma}I_{2j+1}(\gamma).
 \label{eq:ajexact}
\end{equation}
The normalized fundamental one-link expectation is
\begin{equation}
 \frac{a_{1/2}(\gamma)}{2a_0(\gamma)}
 =\frac{I_2(\gamma)}{I_1(\gamma)},
 \label{eq:afundratio}
\end{equation}
which is Eq.~\eqref{eq:tBessel}.  No sum index $i$ is used here or elsewhere; the representation label is $j$.

The integral
\begin{equation}
 z(\gamma)=\int dW\,e^{\frac{\gamma}{2}\operatorname{tr}W}
 \label{eq:zint}
\end{equation}
can be evaluated by writing $W=\cos\alpha+i\hat n\cdot\bm\sigma\sin\alpha$ and using the Haar measure on $S^3$.  Differentiation gives
\begin{equation}
 \frac{d\log z}{d\gamma}
 =\left\langle\frac12\operatorname{tr}W\right\rangle
 =\frac{I_2(\gamma)}{I_1(\gamma)}.
 \label{eq:zdiff}
\end{equation}

The perturbative character expansion at fixed $\beta=0$ and the hopping
series are finite-volume, fixed-cutoff statements. For the leading
path contribution we assume a fixed finite lattice, finite radial moments,
and a specified non-self-intersecting open path; the minimum-length term
then has controlled powers of $\kappa$, while detours, contact terms,
and closed-loop attachments enter at higher (geometry-dependent) orders.
Taking $L\to\infty$ or $a\to0$ term by term is not justified solely
by this lowest-order coefficient.

\subsection{Hopping expansion with a radial mode}

At small $\kappa$, expand each link Boltzmann factor in Eq.~\eqref{eq:latactionrad}.  Gauge integration forces fundamental matrices to form connected paths and closed loops.  For a two-point function with a specified reference dressing, the leading contribution is an open scalar path closed by the reference transporter.  Each internal site supplies a radial moment and each hop supplies a group-integration factor.  This yields Eq.~\eqref{eq:hopPsi}.
\begin{equation}
 m_n=\frac{\int_0^\infty d\rho\,\rho^{n+3}e^{-V(\rho)}}
 {\int_0^\infty d\rho\,\rho^3e^{-V(\rho)}}.
 \label{eq:mn}
\end{equation}
\begin{align}
 C_{\Psi}(x,y) \simeq
 \sum_{\Gamma:x\to y}
 \left(\frac{\kappa}{4}\right)^{L(\Gamma)}
 m_2^{L(\Gamma)+1}
 \bigl\langle W_F(C_{\Gamma})\bigr\rangle_{\rm YM}
 +\cdots.
 \label{eq:hopPsi}
\end{align}
For a Gaussian radial potential $V(\rho)=M_0^2\rho^2$, the moments follow from
\begin{equation}
 \int_0^{\infty}d\rho\,\rho^{p}e^{-M_0^2\rho^2}
 =\frac12 M_0^{-p-1}\Gamma\left(\frac{p+1}{2}\right),
 \label{eq:gaussianmoment}
\end{equation}
therefore in particular $m_2=2/M_0^2$.

For a path $\Gamma$ with $L$ links one must specify whether internal
vertices are revisited and whether the dressing path closes a simple
loop; otherwise the radial moment combinatorics can change.

The moment identity and leading coefficient are exact under the stated
Gaussian potential, but an $O(\kappa^{L+2})$ remainder assumes the path
geometry and finite-lattice hopping expansion are fixed. It should not
be used as a uniform-in-$L$ or continuum error bound.

\subsection{Exact radial integration at $\beta=0$}

At $\beta=0$ change variables from $U_{x,\mu}$ to the scalar-frame link $\widetilde W_{x,\mu}$.  Haar invariance makes the Jacobian unity for the angular change, while the scalar site measure remains $\rho_x^3d\rho_x$.  Integrating each link gives the radial partition function
\begin{equation}
 Z_{\beta=0}=\int\prod_x\left[\rho_x^3d\rho_x e^{-V(\rho_x)}\right]
 \prod_{x,\mu}z(\kappa\rho_x\rho_{x+\hat\mu}).
 \label{eq:Zradbeta0}
\end{equation}

For the normalized radial measure defined by
Eq.~\eqref{eq:Zradbeta0}, the exact conditional Wilson-tree
orientation correlator is
\begin{align}
 C_q^{(T)}(x,y)&=\left\langle
 \prod_{\ell\in T_{xy}}
 \frac{I_2(\kappa\rho_{s(\ell)}\rho_{t(\ell)})}
 {I_1(\kappa\rho_{s(\ell)}\rho_{t(\ell)})}
 \right\rangle_{\!\rm rad}.
 \label{eq:Cqradexact}
\end{align}
Insertion of a fundamental orientation matrix on a tree link replaces $z(s)$ by $z(s)t(s)$.  Therefore the product formula Eq.~\eqref{eq:Cqradexact} follows.

To make the one-link calculation explicit, parametrize
$W=\cos\theta\,\bm1+i\sin\theta\,\widehat n^A\sigma^A$.
The normalized Haar class measure gives
\begin{equation}
 z(s)=\frac2\pi\int_0^\pi d\theta\,
 \sin^2\theta\,e^{s\cos\theta},
 \qquad \langle W\rangle_s=\frac{z'(s)}{z(s)}\bm1.
 \label{eq:haarbesseladd}
\end{equation}
The independent zero-hopping moments $m_1$ and $m_2$
are those of Eq.~\eqref{eq:mn}, evaluated in the normalized
$\kappa=0$ single-site radial measure.
For a simple tree path of $L\ge1$ distinct links, the
orientation-only and unnormalized correlators have different
leading radial moments:
\begin{align}
 C_q^{(T)}(x,y)&=
 \left(\frac\kappa4\right)^L m_1^2m_2^{L-1}
 +O(\kappa^{L+2}),
 \label{eq:mqsmalladd}\\
 C_\Psi^{(T)}(x,y)&=
 \left(\frac\kappa4\right)^L m_2^{L+1}
 +O(\kappa^{L+2}).
 \label{eq:mpsmalladd}
\end{align}
These statements assume sufficiently finite radial moments and
fixed finite volume before expansion. Backtracking paths and
self-intersections have different moment combinatorics.

The small-$s$ expansion of the normalized group integral, used in
the radial effective measure, is
\begin{equation}
 z(s)=1+\frac{s^2}{8}+\frac{s^4}{192}+O(s^6).
 \label{eq:zsmallsadd}
\end{equation}
This expansion distinguishes corrections to the radial measure
from corrections to the normalized conditional link mean $t(s)$.

For $s\ll1$,
\begin{equation}
 t(s)=\frac{s}{4}-\frac{s^3}{96}+\frac{s^5}{1536}+O(s^7),
 \label{eq:tsradsmall}
\end{equation}
whereas for $s\gg1$,
\begin{equation}
 t(s)=1-\frac{3}{2s}+\frac{3}{8s^2}+O(s^{-3}).
 \label{eq:tsradlarge}
\end{equation}
If the radial distribution is sharply localized near $v$, a path of length $L$ has
\begin{equation}
 C_q^{(T)}(L)\simeq
 \exp\left[-\frac{3L}{2\kappa v^2}+O((\kappa v^2)^{-2})\right],
 \qquad
 \xi_T\simeq\frac{2\kappa v^2}{3}.
 \label{eq:xitree}
\end{equation}
This explicitly shows how the radial amplitude controls the reference-frame stiffness.

\subsection{Abelian comparison}

For a fixed-length compact $U(1)$ scalar at $\beta=0$ the one-link integral is $I_0(s)$ and the orientation correlator is controlled by
\begin{equation}
 t_{U(1)}(s)=\frac{I_1(s)}{I_0(s)}.
 \label{eq:U1t}
\end{equation}
The parallel with the $SU(2)$ Bessel ratio makes clear that the exact solvable point tests the relational-frame mechanism rather than a peculiarity of the non-Abelian group.

\section{Reduction-field expansion and lattice Green functions}
\label{qrfA:app:deriv}

At $\beta=0$, the scalar-frame links
$W_{x,\mu}=\hat\Theta_x^\dagger U_{x,\mu}
\hat\Theta_{x+\hat\mu}$ are independent, with normalized weight
\begin{align}
 dP_\gamma(W)=&\frac{dW}{z(\gamma)}
 \exp\!\left[\frac{\gamma}{2}\operatorname{tr}W\right],
 \nonumber\\
  z(\gamma)
=:& \int_{SU(2)} dW \exp\!\left[\frac{\gamma}{2}\operatorname{tr}W\right].
 \label{qrfA:eq:measure}
\end{align}
Writing a single link variable $W$ in $SU(2)$  as
\begin{equation}
 W
= e^{i \theta\,\widehat n^A\sigma_A}  =\cos\theta+i\sin\theta\,\widehat n^A\sigma_A \in SU(2),
 \label{eq:param}
\end{equation}
we have
\begin{equation}
 {1\over2}\mathrm{tr}\,W=\cos\theta,
 \qquad
 dW={2\over\pi}\sin^2\theta\,d\theta\,d\Omega_{\widehat n} .
 \label{eq:haar}
\end{equation}
Therefore, one-link partition function, with the angular part $\Omega_{\widehat n}$ normalized, is given by
\begin{equation}
 z(\gamma)
 ={2\over\pi}\int_0^\pi d\theta\,\sin^2\theta\,e^{\gamma\cos\theta}
 ={2I_1(\gamma)\over\gamma} .
 \label{eq:z}
\end{equation}

\subsection{Variation of the CDG reduction condition}

The CDG reduction functional (\ref{eq:reduce-original}) is rewritten in terms of $\bm r_x$ and $W_{x,\mu}$ into
\begin{align}
 \min_{|\bm r_x|=1}&\mathcal R[\bm r;W], \nonumber\\
 \mathcal R[\bm r;W]
 =&\sum_{x,\mu} [1-\bm r_x\cdot R(W_{x,\mu})\bm r_{x+\hat\mu}] \nonumber\\
 =&\frac12 \sum_{x,\mu} [\bm r_x- R(W_{x,\mu})\bm r_{x+\hat\mu}]^2,\quad |\bm r_x|=1 ,
 \label{qrfA:eq:reduce}
\end{align}
where $R(W)$ is the adjoint rotation:
\begin{align}
 R(W_{x,\mu})\bm r_{x+\hat\mu}:= W_{x,\mu} \bm r_{x+\hat\mu} W_{x,\mu}^\dagger ,
 \label{adj-rot}
\end{align}

Writing $R_{x,\mu}=R(W_{x,\mu})$, introduce site multipliers
$\eta_x$ to enforce $\bm r_x^2=1$:
\begin{equation}
 \mathcal F=\mathcal R[\bm r;W]
 +\frac12\sum_x\eta_x(\bm r_x^2-1).
 \label{qrfA:eq:F}
\end{equation}
The direction $\bm r_x$ occurs in outgoing and incoming terms.
Consequently,
\begin{align}
 \frac{\partial\mathcal F}{\partial\bm r_x}
 &=-\bm H_x+\eta_x\bm r_x=0,
 \nonumber\\
 \bm H_x&=\sum_{\mu=1}^4
 \bigl[R_{x,\mu}\bm r_{x+\hat\mu}
 +R_{x-\hat\mu,\mu}^{T}\bm r_{x-\hat\mu}\bigr].
 \label{qrfA:eq:stat}
\end{align}
The constrained stationarity equation is
\begin{equation}
 \bm r_x\times\bm H_x=0,
 \quad |\bm r_x|=1.
 \label{qrfA:eq:stationary}
\end{equation}
A stationary solution can be a local minimum or another stationary
configuration; neither the iteration nor the harmonic expansion
below ensures the absolute reduction minimum or a unique copy.

For $\gamma\gg1$ set
\begin{equation}
W_{x,\mu}=\exp(\mathrm i\bm a_{x,\mu}\cdot\bm\sigma),\qquad
\bm a_{x,\mu}=O(\gamma^{-1/2}).
\label{eq:a}
\end{equation}
Expand the adjoint rotation as
\begin{align}
R(W)\bm v=&W \bm v W^\dagger
= \exp(\mathrm i\bm a\cdot\bm\sigma) \bm v \exp(-\mathrm i\bm a\cdot\bm\sigma) \nonumber\\
=&\bm v-2\bm a\times\bm v
+2\bm a\times(\bm a\times\bm v)+O(|\bm a|^3).
\label{eq:rotationexp}
\end{align}

With respect to a fixed unit vector $\bm n_0$, any vector in its neighborhood admits the orthogonal decomposition
\begin{equation}
\bm r_x=\alpha_x\bm n_0+\bm\pi_x,
\qquad \bm n_0\cdot\bm\pi_x=0
\label{eq:decomposition}
\end{equation}
From $|\bm r_x|^2=1$ it follows that
$\alpha_x^2+|\bm\pi_x|^2=1$.
Choosing the branch near $\bm n_0$ ($\alpha_x>0$), we obtain
\begin{equation}
 \bm r_x=\sqrt{1-|\bm\pi_x|^2}\,\bm n_0+\bm\pi_x,\qquad
|\bm\pi_x|<1.
\label{eq:exactcoord}
\end{equation}
This is an exact identity locally. Expanding the square root gives
\begin{align}
\bm r_x=&\left(1-\frac12|\bm\pi_x|^2
-\frac18|\bm\pi_x|^4+\cdots\right)\bm n_0+\bm\pi_x \nonumber\\
=&\bm n_0+\bm\pi_x-\tfrac12|\bm\pi_x|^2\bm n_0+O(|\bm\pi_x|^3).
\label{eq:taylorr}
\end{align}
In particular, if these coordinates are adopted exactly, the next term in the geometrical expansion is of order $O(|\bm\pi_x|^4)$.
The $O(|\bm\pi_x|^3)$ remainder is correct, but provides a weaker estimate of the error.

The linear part of $\bm r_x-R(W)\bm r_{x+\hat\mu}$ is
\begin{equation}
\bm r_x-R(W_{x,\mu})\bm r_{x+\hat\mu}
=-\left(\bm\pi_{x+\hat\mu}-\bm\pi_x-2\bm a_{x,\mu}\times\bm n_0\right)+O(a^2).
\label{eq:difference}
\end{equation}
Therefore the functional in the harmonic approximation reads
\begin{equation}
\mathcal R^{(2)}[\bm\pi;\bm a]=\frac12\sum_{x,\mu}
\left|\Delta_\mu\bm\pi_x-\bm c_{x,\mu}\right|^2,
\quad \bm c_{x,\mu}:=2\bm a_{x,\mu}\times\bm n_0,
\label{eq:harmonic}
\end{equation}
where
\begin{equation}
\Delta_\mu f_x=f_{x+\hat\mu}-f_x.
\label{eq:forwarddifference}
\end{equation}
The overall choice of $\bm n_0$ is not unique at this order. The analysis below is a harmonic expansion around a local branch with $\bm n_0$ fixed. The effects of optimizing $\bm n_0$ configuration by configuration in the global problem must be reconsidered, especially at finite volume and higher orders.

Under $\bm\pi_x\to\bm\pi_x+\delta\bm\pi_x$,
\begin{align}
&\delta\mathcal R^{(2)}
 =\sum_{x,\mu}(\Delta_\mu\bm\pi_x-\bm c_{x,\mu})
 \cdot\Delta_\mu\delta\bm\pi_x
 \\
&=\sum_x\delta\bm\pi_x\cdot
\sum_\mu\left[2\bm\pi_x-\bm\pi_{x+\hat\mu}-\bm\pi_{x-\hat\mu}
+\bm c_{x,\mu}-\bm c_{x-\hat\mu,\mu}\right]
 \\
&= \sum_x\delta\bm\pi_x\cdot
 \bigl[(-\Delta_{\rm lat})\bm\pi_x
 +(\nabla^-\!\cdot\bm c)_x\bigr]=0.
  \label{qrfA:eq:variation}
\end{align}
Here we used discrete integration by parts on the periodic lattice, equivalently relabeling the site indices. Define the lattice Laplacian and backward divergence by
\begin{align}
 (-\Delta_{\rm lat})f_x
 &=\sum_{\mu=1}^4(2f_x-f_{x+\hat\mu}-f_{x-\hat\mu}),
 \nonumber\\
 (\nabla^-\!\cdot\bm c)_x
 &=\sum_\mu(\bm c_{x,\mu}-\bm c_{x-\hat\mu,\mu}).
 \label{qrfA:eq:lap}
\end{align}

The stationarity equation can then be written as
\begin{equation}
(-\Delta_{\rm lat})\bm\pi_x=-(\nabla^-\!\cdot\bm c)_x .
\label{eq:poisson}
\end{equation}
Thus the harmonic minimizer satisfies the \emph{lattice Poisson
equation}
\begin{align}
 (-\Delta_{\rm lat})\bm\pi_x
 &=-2\sum_\mu
 (\bm a_{x,\mu}-\bm a_{x-\hat\mu,\mu})
 \times\bm n_0.
 \label{qrfA:eq:Poisson}
\end{align}
The summed source vanishes identically. Choosing
$\sum_x\bm\pi_x=0$ removes the zero mode and fixes the solution
of this \emph{linearized} problem. This does not settle the
optimization over $\bm n_0$ or possible reduction copies.
The right-hand side sums to zero over all sites and therefore has no projection onto the zero mode. Choosing $\sum_x\bm\pi_x=0$ makes the linear solution unique; this is not a statement about uniqueness of the original nonlinear reduction problem.

\subsection{Fourier inversion and the $8/\gamma$ coefficient}

Let $V$ be the volume $V=L^4$ of the lattice with the size $L$.
Define
\begin{equation}
k_\mu=2\pi m_\mu/L ,  \quad
m_\mu=0,\ldots,L-1 .
\end{equation}
 Then the Fourier transform is defined by
\begin{equation}
f_x:=\frac1{\sqrt V}\sum_k e^{\mathrm ik\cdot x}\widetilde f(k),\quad
\widehat k^{\,2}:=4\sum_{\mu=1}^4\sin^2(k_\mu/2),\quad
\label{eq:fourier}
\end{equation}
Since
$
(-\Delta_{\rm lat})e^{\mathrm ikx}=\widehat k^{\,2}e^{\mathrm ikx}
$
 and
$\widetilde{\nabla^-\!\cdot\bm c}(k)
=\sum_\mu(1-e^{-ik_\mu})\widetilde{\bm c}_\mu(k)$,
Eq.~\eqref{qrfA:eq:Poisson} gives, for $k\ne0$,
\begin{equation}
 \widetilde{\bm\pi}(k)
 =-\frac{\sum_\mu(1-e^{-ik_\mu})
 \widetilde{\bm c}_\mu(k)}{\widehat k^{\,2}}.
 \label{qrfA:eq:pisol}
\end{equation}
For $W=\exp(\mathrm i\bm a\cdot\bm\sigma)$,
\begin{equation}
 \tfrac\gamma2\operatorname{tr}W=\gamma\cos|\bm a|
=\gamma-\tfrac\gamma2|\bm a|^2+O(\gamma |\bm a|^4).
 \label{qrfA:eq:acov}
\end{equation}

At $\beta=0$, expand around the identity with
$W_{x,\mu}=\exp(i\bm a_{x,\mu}\cdot\bm\sigma)$.
The \emph{normalized} leading Gaussian one-link probability measure is
\begin{equation*}
 d\mu_G(\bm a)=\left(\frac{\gamma}{2\pi}\right)^{3/2}
 \exp\!\left(-\frac\gamma2|\bm a|^2\right)d^3a,
 \quad \int_{\mathbb R^3}d\mu_G(\bm a)=1.
\end{equation*}
For independent links, this gives the Gaussian covariance
\begin{equation}
 \langle a_{x,\mu}^A a_{y,\nu}^B\rangle_G
 =\frac1\gamma\delta^{AB}\delta_{xy}\delta_{\mu\nu}.
 \label{eq:alinkcov}
\end{equation}
The normalized exact $SU(2)$ one-link measure, including its Haar
Jacobian and non-Gaussian terms, instead has the asymptotic covariance
$\langle a_{x,\mu}^A a_{y,\nu}^B\rangle_0
=\gamma^{-1}\delta^{AB}\delta_{xy}\delta_{\mu\nu}
+O(\gamma^{-2})$ at fixed links, in the local identity chart.
All following $O(\gamma^{-1})$ harmonic covariances use
$\langle\cdot\rangle_G$; replacing it by the exact normalized
one-link expectation changes the quoted leading terms only at higher order.

For an arbitrary fixed unit color vector
$\bm n_0=(n_0^1,n_0^2,n_0^3)$ with $|\bm n_0|^2=1$,
introduce the orthogonal projector onto its tangent plane,
$P_\perp^{AB}(\bm n_0):=\delta^{AB}-n_0^A n_0^B$
($A,B=1,2,3$).  It obeys
$P_\perp^{AB}n_0^B=0$,
$P_\perp^{AC}P_\perp^{CB}=P_\perp^{AB}$, and
$\sum_{A=1}^3P_\perp^{AA}=2$.
Since $c_{x,\mu}^A=2\epsilon^{ABC}a_{x,\mu}^B n_0^C$,
the normalized Gaussian covariance in Eq.~\eqref{eq:alinkcov} and the identity
$\epsilon^{ACD}\epsilon^{BCE}n_0^D n_0^E
=\delta^{AB}-n_0^A n_0^B$ give
\begin{equation}
\left\langle\widetilde c_\mu^A(k)
\widetilde c_\nu^B(-k)\right\rangle_G
=\frac4\gamma P_\perp^{AB}(\bm n_0)\delta_{\mu\nu}.
\label{eq:ccov-arbitrary-n0}
\end{equation}
This is valid in any fixed color coordinate system; no choice
$\bm n_0=(0,0,1)$ is needed.
Substituting into Eq.~\eqref{qrfA:eq:pisol}, for $k\ne0$,
\begin{align}
\left\langle\widetilde\pi^A(k)
\widetilde\pi^B(-k)\right\rangle_G
&=\frac{4P_\perp^{AB}(\bm n_0)}{\gamma}
\frac{\sum_\mu|1-e^{-\mathrm ik_\mu}|^2}
{(\widehat k^{\,2})^2}\nonumber\\
&=\frac{4P_\perp^{AB}(\bm n_0)}
{\gamma\widehat k^{\,2}},\qquad k\ne0.
\label{qrfA:eq:picov}
\end{align}
Here $|1-e^{-\mathrm ik_\mu}|^2=4\sin^2(k_\mu/2)$
and $\sum_\mu|1-e^{-\mathrm ik_\mu}|^2=\widehat k^{\,2}$.

Introducing the zero-mode-subtracted Green function
\begin{equation}
G_L(x-y):=\frac1{L^4}\sum_{k\ne0}
\frac{e^{\mathrm ik\cdot(x-y)}}{\widehat k^{\,2}}.
\label{eq:Gdef}
\end{equation}
we obtain the color-covariant position-space covariance,
\begin{equation}
\left\langle\pi_x^A\pi_y^B\right\rangle_G
=\frac4\gamma P_\perp^{AB}(\bm n_0)G_L(x-y),
\qquad A,B=1,2,3.
\label{eq:picov-position-arbitrary-n0}
\end{equation}
Taking the trace over all three Cartesian color components,
which automatically counts only the two transverse directions,
\begin{align}
\left\langle|\bm\pi_x-\bm\pi_y|^2\right\rangle_G
&=\frac8\gamma\sum_{A=1}^3
P_\perp^{AA}(\bm n_0)
\,[G_L(0)-G_L(x-y)]\nonumber\\
&=\frac{16}{\gamma}
[G_L(0)-G_L(x-y)].
\label{eq:pidiff}
\end{align}
Finally, the local expansion
$\bm r_x\cdot\bm r_y
=1-\frac12|\bm\pi_x-\bm\pi_y|^2+O(\bm\pi^4)$
gives
\begin{equation}
C_n(r)=\left\langle\bm r_x\cdot\bm r_y\right\rangle_G
=1-\frac8\gamma[G_L(0)-G_L(r)]+O(\gamma^{-2}).
\label{eq:Cleading}
\end{equation}
Here $r=0,1,\ldots,\lfloor L/2\rfloor$ denotes separation
along a lattice axis.  The coefficient $8/\gamma$ is
independent of the orientation of the fixed unit vector
$\bm n_0$, because $\operatorname{tr}P_\perp=2$.
This is a large-$\gamma$ asymptotic result on a fixed local
harmonic branch at finite $L$; it does not establish the
same covariance after choosing $\bm n_0$ adaptively from
the gauge-link configuration, nor does it prove a global
reduction minimum or uniqueness of the reduction copy.

\subsection{Explicit finite-$L$ and infinite-volume Green functions}

In infinite volume the Schwinger representation gives
\begin{align}
 G_\infty(r)&=\int_{[-\pi,\pi]^4}
 \frac{d^4k}{(2\pi)^4}
 \frac{\cos(k_1r)}{\widehat k^{\,2}}
 \nonumber\\
 &=\int_0^\infty dt\,e^{-8t}I_r(2t)I_0(2t)^3.
 \label{qrfA:eq:GSchwinger}
\end{align}
Alternatively, integrating over $k_1$ first yields
\begin{align}
 G_\infty(r)
 &=\int_{[-\pi,\pi]^3}\frac{d^3p}{(2\pi)^3}
 \frac{e^{-r\omega(\bm p)}}{2\sinh\omega(\bm p)},
 \nonumber\\
 \cosh\omega(\bm p)
 &=1+2\sum_{j=2}^4\sin^2(p_j/2).
 \label{qrfA:eq:G3}
\end{align}
Using $\omega(\bm p)\simeq|\bm p|$ as $\bm p\to0$,
\begin{equation}
 G_\infty(r)=\frac1{4\pi^2r^2}+O(r^{-4}),
 \quad \lim_{r\to\infty}G_\infty(r)=0.
 \label{qrfA:eq:long}
\end{equation}
Numerical integration gives
\begin{equation}
8G_\infty(0)=1.239467121849\ldots .
\end{equation}

\begin{table}[th]
\caption{Infinite-volume Green functions and coefficients
 $a_r=8[G_\infty(0)-G_\infty(r)]$; numerical integration of
 Eq.~\eqref{qrfA:eq:GSchwinger}.}
\label{qrfA:tab:G}
\centering\begin{tabular}{rcc}\hline\hline
$r$ & $G_\infty(r)$ & $a_r$ \\ \hline
0&0.154933390231&0\\
1&0.029933390231&1.000000000000\\
2&0.008245508992&1.173503049914\\
3&0.003286704243&1.213173487905\\
4&0.001721255961&1.225697074158\\
8&0.000402433463&1.236247654147\\
16&0.000099339241&1.238672407918\\
$\infty$&0&1.239467121849\\ \hline\hline
\end{tabular}
\end{table}

\begin{figure}[th]
 \centering
 \includegraphics[width=.92\columnwidth]{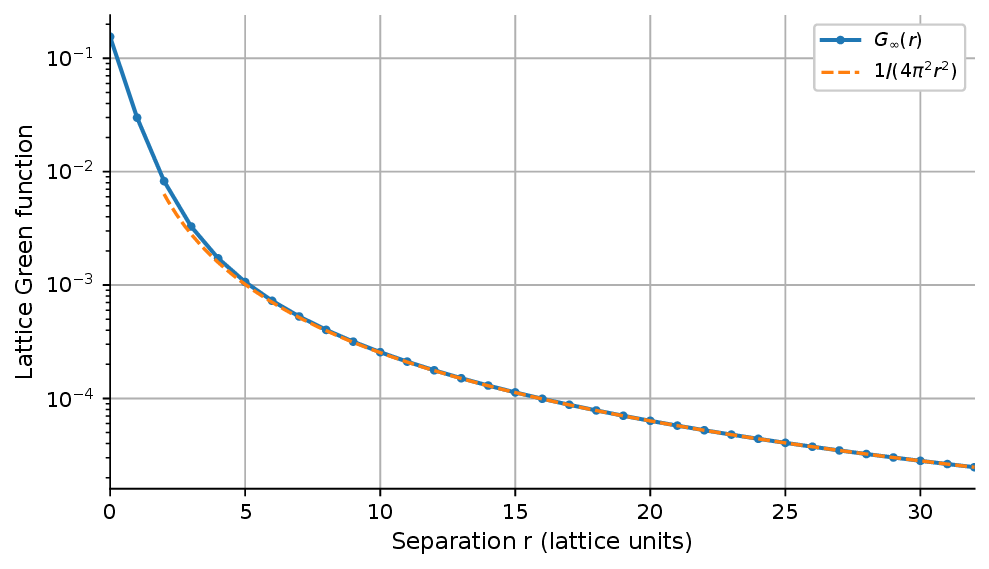}
 \caption{The four-dimensional infinite-volume lattice Green
 function $G_\infty(r)$ and the $1/(4\pi^2r^2)$ asymptote.}
 \label{qrfA:fig:green}
\end{figure}

\begin{figure}[th]
 \centering
 \includegraphics[width=.92\columnwidth]{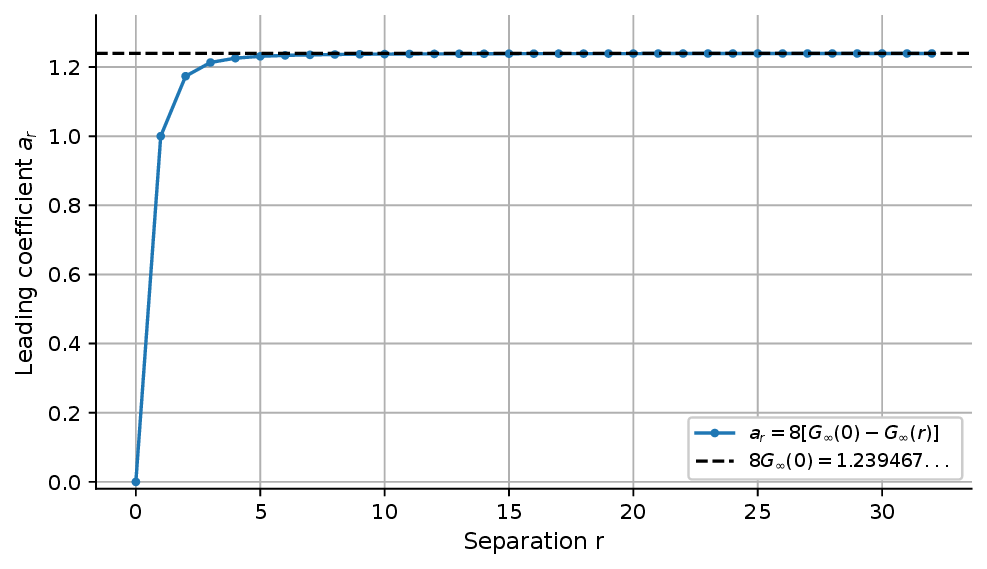}
 \caption{The coefficient $a_r$ approaches
 $8G_\infty(0)=1.239467\ldots$ as $r\to\infty$.}
 \label{qrfA:fig:ar}
\end{figure}

For $\bm m_\perp=(m_2,m_3,m_4)$ define
\begin{equation}
 M^2(\bm m_\perp)=4\sum_{\mu=2}^4
 \sin^2(\pi m_\mu/L),\qquad
 \cosh\omega=1+\frac{M^2}{2}.
 \label{qrfA:eq:M}
\end{equation}
The $k_1$ sum can be performed for $M>0$:
\begin{align}
 S_L(r;M^2)
 &=\frac1L\sum_{m_1=0}^{L-1}
 \frac{\cos(2\pi m_1r/L)}
 {M^2+4\sin^2(\pi m_1/L)}
 \nonumber\\
 &=\frac{\cosh[\omega(L/2-r)]}
 {2\sinh\omega\,\sinh(\omega L/2)}.
 \label{qrfA:eq:SL}
\end{align}
This follows by summing the infinite-line propagator
$e^{-\omega|r|}/(2\sinh\omega)$ over periodic images.
For $\bm m_\perp=0$ remove $m_1=0$ and obtain
\begin{equation}
 S_L^{(0)}(r)=\frac{L^2-1}{12L}
 -\frac{r(L-r)}{2L}.
 \label{qrfA:eq:S0}
\end{equation}
Combining the modes gives an explicit threefold sum:
\begin{align}
 G_L(r)=\frac1{L^3}\Biggl\{
 &\frac{L^2-1}{12L}-\frac{r(L-r)}{2L}
 \nonumber\\
 &+\sum_{\bm m_\perp\ne0}
 \frac{\cosh[\omega_{\bm m_\perp}(L/2-r)]}
 {2\sinh\omega_{\bm m_\perp}
 \sinh(\omega_{\bm m_\perp}L/2)}\Biggr\}.
 \label{qrfA:eq:Gexplicit}
\end{align}
It is periodic, $G_L(r+L)=G_L(r)$; hence there is no literal
$r\to\infty$ limit at fixed $L$. The zero-mode-subtracted
Green function satisfies
\begin{align}
 (-\Delta_{\rm lat})G_L(x)&=\delta_{x,0}-L^{-4},\nonumber\\
 8[G_L(0)-G_L(1)]&=1-L^{-4}.
 \label{qrfA:eq:ident}
\end{align}

For the independent-link measure, group character orthogonality gives
\begin{align}
 \lambda_1(\gamma)
 &=\frac{1}{3z(\gamma)}\int dW\,
 e^{\frac\gamma2\operatorname{tr}W}\chi_1(W)
 \nonumber\\
 &=\frac{I_2(\gamma)-I_4(\gamma)}{3z(\gamma)}
 =\frac{I_3(\gamma)}{I_1(\gamma)}.
 \label{qrfA:eq:lambdader}
\end{align}
Here $\chi_1(W)=1+2\cos(2\vartheta)$ for
$W=\cos\vartheta+i\sin\vartheta\hat{\bm a}\cdot\bm\sigma$.
The transported CDG coefficient in the original small-$\gamma$
proposal is
\begin{equation}
 K_r^{(0)}=\Big\langle\bm n_{\mathscr A,x}
 \cdot R(U_{P_{xy}})\bm n_{\mathscr A,y}\Big\rangle_{\gamma=0},
 \label{qrfA:eq:K}
\end{equation}
with $P_{xy}$ a fixed length-$r$ straight path.

The conditional short-distance expression
$C_n(r)\simeq K_r^{(0)}[I_3(\gamma)/I_1(\gamma)]^r$
uses the exact one-link transfer ratio, while the coefficient
$K_r^{(0)}$ depends on the separately specified reduction protocol.
It is a limited hopping estimate, not an all-distance identity or a numerical comparison.

\end{document}